\documentclass[twocolumn]{aastex63}
\usepackage{apjfonts}
\usepackage{amsmath}
\usepackage{multirow}
\usepackage{graphics}
\usepackage{threeparttable}
\usepackage{hyperref}
\usepackage{lineno}
\usepackage{soul}

\shorttitle{MW potential from snail intersections}
\shortauthors{Guo et al.}
\begin{document}

\title{Measuring the Milky Way Vertical Potential with the Phase Snail in a Model Independent Way}

\author{Rui Guo}%% [] for orcid

%The affiliation
\affiliation{Department of Astronomy, School of Physics and Astronomy, Shanghai Jiao Tong University, 800 Dongchuan Road, Shanghai 200240, China}
\affiliation{Key Laboratory for Particle Astrophysics and Cosmology (MOE) / Shanghai Key Laboratory for Particle Physics and Cosmology, Shanghai 200240, China}

\author{Zhao-Yu Li}
\affiliation{Department of Astronomy, School of Physics and Astronomy, Shanghai Jiao Tong University, 800 Dongchuan Road, Shanghai 200240, China}
\affiliation{Key Laboratory for Particle Astrophysics and Cosmology (MOE) / Shanghai Key Laboratory for Particle Physics and Cosmology, Shanghai 200240, China}

\author{Juntai Shen}
\affiliation{Department of Astronomy, School of Physics and Astronomy, Shanghai Jiao Tong University, 800 Dongchuan Road, Shanghai 200240, China}
%%\email{jtshen@sjtu.edu.cn; lizy.astro@sjtu.edu.cn}
%%\correspondingauthor{jtshen@sjtu.edu.cn; lizy.astro@sjtu.edu.cn}
\affiliation{Key Laboratory for Particle Astrophysics and Cosmology (MOE) / Shanghai Key Laboratory for Particle Physics and Cosmology, Shanghai 200240, China}
%%\affiliation{Shanghai Astronomical Observatory, Chinese Academy of Sciences, 80 Nandan Road, Shanghai 200030, China}

\author{Shude Mao}
\affiliation{Department of Astronomy, Tsinghua University, Beijing 100084, China}

\author{Chao Liu}
\affiliation{Key Lab of Space Astronomy and Technology, National Astronomical Observatories, Chinese Academy of Sciences, 20A Datun Road, Chaoyang District, Beijing 100101, China}

% \correspondingauthor{Rui Guo}
% \email{guorui20@sjtu.edu.cn}
\correspondingauthor{Zhao-Yu Li, Juntai Shen}
\email{lizy.astro@sjtu.edu.cn; jtshen@sjtu.edu.cn}

%



\begin{abstract}
The vertical phase-space spiral (snail) is a direct sign of dis-equilibrium of Milky Way's disc. Nevertheless, the wrapping of the phase snail contains information of the vertical potential. We propose a novel method to measure the vertical potential utilizing the intersections between the snail and $z$/$V_{z}$ axes, for which we know the maximum vertical heights ($Z_{max}$) or the maximum vertical velocities ($V_{z,max}$). Using a refined linear interpolation method, we directly obtain $(Z_{max},\ \frac{1}{2}V_{z,max}^{2})$ for these snail intersections to constrain the vertical potential profile empirically. Our method is model independent since no assumptions about the snail shape or the vertical potential have been made. Although the snail binned by the guiding center radius ($R_{g}$) is more prominent, it traces a vertical potential shallower than that of the snail binned by the same Galactocentric radius ($R$). We apply an empirical method to correct this difference. We measure the snail intersections in several $R_{g}$ bins within $7.5< R_{g} < 11.0$ kpc for Gaia DR3, and apply the interpolation method to deduce the potential values at several vertical heights. The potential at the snail intersections, as well as the following mass modeling are consistent with the popular Milky Way potentials in the literature. For the $R_{g}$-binned phase snail in the Solar neighborhood, the mass modeling indicates a local dark matter density of $\rho_{\rm dm}= 0.0150\pm0.0031$ $\rm M_{\odot}\,pc^{-3}$, consistent with previous works. Our method could be applied to larger radial ranges in future works, to provide independent and stronger constraints on the Milky Way's potential.

\end{abstract}



%%%%%%%%%%%%%%%%%%%%%%%%%%%%%%%%%%%%%%%%%%%%%%%%%%%%%%%%%%%%%%
\section{Introduction} 
\label{sec:intro}
A main goal of dynamical studies of our Milky Way is to investigate the mass distribution, consisting of a thin disk, thick disk, bulge/bar, stellar halo, and a dark matter halo \citep{Bland-Hawthorn2016}. The different characteristics and interlinks of the various Galactic components could reveal the formation and evolution history of our Milky Way \citep[e.g.][]{Read2008, Helmi2018, Mackereth2019, Helmi2020}. For example, the shape of the Galactic dark matter halo helps to understand the influence of the baryonic effects to the structure formation, and thus the current galaxy formation models \citep[e.g.][]{Kazantzidis2010, Koposov2010, Lux2012, Eadie2019, Fardal2019, Iorio2019, Wegg2019}. In the Solar neighborhood, many dynamical studies aim to measure the local dark matter density, which is an important physical parameter to provide guidance for the direct dark matter detection experiments \citep[e.g.][]{Fairbairn2013, Green2017, Maity2021, Bechtol2022}.

Various methods have been applied to measure the Galactic potential. In the local region, the vertical potential and mass distribution have been estimated by analyzing the kinematics of different stellar tracers, employing the vertical Jeans equation or the distribution function method \citep[e.g.][]{ZhangL2013, XiaQR2016, Hagen2018, Schutz2018, Sivertsson2018, Buch2019, GuoR2020, Salomon2020}. The local baryonic census studies help to clarify the baryonic contribution to the Galactic potential and provide appropriate priors to the modeling \citep[e.g.][]{Holmberg2000, Flynn2006, McKee2015}. Some of more global studies of the Galactic mass distribution usually fit a specific galactic mass model to the Galactic rotation curve, which is obtained by the compilation of kinematic measurements of gas, stars and masers \citep[e.g.][]{Pato2015, HuangY2016, Benito2019, Benito2021, deSalas2019, Karukes2019, LinHN2019, Sofue2020}, or by the radial Jeans equation \citep[e.g.][]{Eilers2019}. According to the recent measurements, the local dark matter density is estimated within $0.4 - 0.6\ {\rm Gev\,cm^{-3} }$ , corresponding to $0.010 - 0.015\ {\rm M_{\odot}\,pc^{-3} }$ , with uncertainties mainly contributed by the observations and the degeneracy between the baryon and dark matter density profiles \citep{Read2014, deSalas2021}.

In other global mass profile studies of Milky Way, tracers such as halo stars, stellar streams, globular clusters and satellites, have also been extensively utilized, especially to study the dark matter halo shape \citep[e.g.][]{Fritz2018, Malhan2019, Hattori2021, Ibata2021, Bird2022}. More recently, the action-angle based distribution function method \citep{Binney2010, Binney2011, McMillan2013} has been employed to provide self-consistent constraints on the Milky Way potential \citep[e.g.][]{Bovy2013, Piffl2014, LiCD2022, Binney2022}. In addition, several Milky Way potential models have been constructed through synthesizing all of the knowledge about the components of the Milky Way into a coherent picture of the gravitational potential \citep[e.g.][]{Irrgang2013, Bovy2015, McMillan2017, Cautun2020}. These potentials have been widely used in the orbit integration and orbital properties estimation, in galpy \citep{Bovy2015} and AGAMA \citep{Vasiliev2019}. (For reviews of local dark matter density measurements, see \citealt{Read2014} and \citealt{deSalas2021}, and for global measurements of Milky Way mass profile, see \citealt{WangWT2020}.)

Most of the dynamical studies are based on the equilibrium assumption, assuming a time invariant distribution function. However, many signs of dis-equilibrium have been found recently in Milky Way. The number density difference between the north and south of the Galactic plane displays a wave-like pattern, showing about 10\% difference at some vertical heights \citep[e.g.][]{Widrow2012, Yanny2013, WangHF2018a, Bennett2019}. The disc shows a non-zero vertical bulk motion, classified into bending and breathing modes in different regions \citep[e.g.][]{Widrow2012, Carlin2013, Carrillo2018, Gaia2018b, WangHF2018b, WangHF2020, Bennett2019}. Recently, \cite{Antoja2018} found clear spiral features in the vertical phase space ($z - V_{z}$) color-coded by the stellar number density, azimuthal velocity and radial velocity, using Gaia DR2 \citep{Gaia2018a}. The phase-space spiral (phase snail) is the result of incomplete phase mixing in the Galactic vertical direction. The phase snail is found to vary with stellar age, chemistry, orbital properties and the location in the Galactic disk (radially and azimuthally) \citep[e.g.][]{TianHJ2018, Bland-Hawthorn2019, Laporte2019, WangC2019, LiZY2020, XuY2020, LiZY2021}. The phase snail has been explained by the passage of a satellite galaxy, such as the last pericentric passage of the Sagittarius dwarf \citep[e.g.][]{Antoja2018, Binney2018, Bland-Hawthorn2019, Laporte2019, Bland-Hawthorn2021}, by the buckling instability of the bar \citep{Khoperskov2019}, by the dark matter wake induced by a satellite passage \citep{Grand2022}, by cumulative effect of smaller intersections with subhalos \citep{Tremaine2023}, or by multiple origins as indicated by the two armed phase spirals in the inner Milky Way \citep{Hunt2022, LiCD2023}.

The observations of the north-south asymmetries and the phase snail contradict the equilibrium assumption in the dynamical modeling. Several works have tried to reconstruct the vertical potential with the consideration of the phase snail. \cite{LiHC2021, LiHC2022} applied steady-state dynamical models to giant stars and found the phase snail in the residual of the best-fit model of the distribution function in the vertical phase space. In a series of works by \cite{Widmark2021a, Widmark2021b, Widmark2022a, Widmark2022b}, the authors predicted the snail feature by the one-dimensional (1D) phase-space orbital integration, and treated it as a component having a constant number density fraction relative to the bulk distribution function. By fitting the 2D phase space distribution in the observation, they estimated the vertical potential embedded in the phase-space orbital integration. In our previous work \citep{GuoR2022}, we applied a geometric snail model superimposed on a smooth background in the phase space to fit the north-south asymmetry of the vertical velocity dispersion profile. Our model successfully explained the north-south asymmetries in the vertical profiles of both the stellar number density and the vertical kinematics. We also obtained an estimation of the background potential which constructs the phase space distribution of the background component through the vertical Jeans equation. Conversely, under a certain Milky Way potential, the phase snail can be utilized to deduce the perturbation time through measuring the slope in the vertical frequency and action angle space \citep[e.g.][]{Darragh-Ford2023, Frankel2022, Widrow2023}.

The phase snail shape curve contains the information of the vertical potential. Once perturbed vertically, stars would oscillate in the vertical direction under the vertical potential, if ignoring the coupling between the vertical motion and the in-plane motion (which is a valid approximation in low vertical regions). The snail then wraps up due to the anharmonic oscillation induced by the vertical potential. In this work, we propose a model independent method to measure the Galactic vertical potential, which utilizes the intersections between the phase snail and the $z/V_{z}$ axes. The intersections have either known maximum vertical heights ($Z_{max}$) along the $z$-axis, or maximum vertical velocities ($V_{z,max}$) along the $V_{z}$-axis. A proper interpolation method can be used to provide accurate ($Z_{max}$, $V_{z,max}$) values for these intersections, which can be used as a direct measurement of the vertical potential values at different vertical heights.

The paper is organized as follows: The method is described and verified with a test particle simulation in Section \ref{sec:method}. In this section, we also study the shape difference between the phase snails binned by the guiding center radius ($R_{g}$) and by the Galactocentric radius ($R$), and propose an empirical correction to account for this difference when measuring the vertical potential using the $R_{g}$-binned phase snails, which is more prominent to be measured accurately. Section \ref{sec:data} describes the observational data used. The measurement of the snail intersections at different $R_{g}$, the comparisons with popular Milky Way potentials and the further mass modeling utilizing the model independent potential $(Z_{max},\ \frac{1}{2}V_{z,max}^{2})$ of the snail intersections are shown in Section \ref{sec:results}. Discussions and conclusions are made in Sections \ref{sec:discuss} and \ref{sec:con}, respectively.

%%%%%%%%%%%%%%%%%%%%%%%%%%%%%%%%%%%%%%%%%%%%%%%%%%%%%%%%%%%%%%
\section{Method and Simulation Verification} 
\label{sec:method}
%%%%%%%%%%%%%%%%%%%%%%%%%%%%%%%%%%%%%%%%%
%%%%%%%%%%%%%%%%%%%%%%%%%%%%%%%%%%%%%%%%%
\subsection{Method description}
\label{ssec:method_desc}
%%%%%%%%%%%%%%%%%%%%%%%%%%%%%%%%%%%%%%%%%
The phase spiral is the result of incomplete phase mixing in the vertical phase space ($z- V_{z}$ space). The phase-space distribution should be steady and smooth for an equilibrium state. However, when the disc is perturbed (externally or internally), stars are disturbed to similar phase angles. As stars have different orbital frequencies, they will gradually wind up to form a spiral structure due to the differential rotation in the phase space. The wrapping of the phase snail and the dynamics behind have been described in great detail in sections 2.1 - 2.3 of \cite{Widmark2021a}. We briefly summarize the picture here and introduce our different method utilized for measuring the Milky Way vertical potential in this work.

The basic assumption of measuring the Milky Way vertical potential with the phase snail is that the vertical and in-plane motions are separable. More specifically, the vertical energy $E_{z}$\footnote{Energy $E_{z}$ here is in fact the energy per unit mass.} is conserved for a vertically oscillating star, which moves in the vertical phase space following a certain closed trajectory with an oscillation frequency ($\Omega_{z}$) and an orbital period ($P(E_{z})$). The vertical energy of a star is equal to
%%%%%%%%%%%%%%%
\begin{equation}
\label{eq:Ez}
    E_{z}= \frac{1}{2} V_{z}^{2}\, +\, \Psi(z) \ ,
\end{equation}
%%%%%%%%%%%%%%%
where the vertical potential energy $\Psi(z)$ is a reduced energy in a Milky Way potential $\Phi(R,z)$ as
%%%%%%%%%%%%%%%
\begin{equation}
\label{eq:Phiz}
    \Psi(z)= \Phi(R,z)\, -\, \Phi(R,0)\ .
\end{equation}
%%%%%%%%%%%%%%%
Given a certain value of $E_{z}$, the phase-space trajectory of a star can be fully determined under certain potential model $\Psi(z)$. The trajectory intersects with $z$ axis at $\pm Z_{max}$ (i.e. the turning-around points with $V_{z}=0$), and with $V_{z}$ axis at $\pm V_{z,max}$ (i.e. the mid-plane points with $z=0$), where the star has the maximum vertical potential and vertical kinetic energy, respectively. Thus, the vertical energy can be related to $Z_{max}$ and $V_{z,max}$ as follows:
%%%%%%%%%%%%%%%
\begin{equation}
\label{eq:Ez_zvzmax}
    E_{z}= \frac{1}{2} V_{z}^{2}\, +\, \Psi(z)= \Psi(Z_{max}) = \frac{1}{2} V_{z,max}^{2}\ .
\end{equation}
%%%%%%%%%%%%%%%
The orbital period $P(E_{z})$ of the star can be calculated by the 1D orbital integration as
%%%%%%%%%%%%%%%
\begin{equation}
\label{eq:Pz}
    P\left(E_{z}\right)= \oint \frac{d z}{\left| V_{z} \right|}= 4\int_0^{Z_{max}} \frac{d z}{\sqrt{2\left[ E_{z}-\Psi(z) \right]}} \ ,
\end{equation}
%%%%%%%%%%%%%%%
where the factor 4 on the right hand side is due to the symmetry about the phase space origin. The vertical frequency is therefore defined as $\Omega_{z}= 2\pi / P\left(E_{z}\right)$. The phase-space trajectory of a star is also determined by a similar integration as Eq. \ref{eq:Pz} but with different upper limits of $|z|$.

\begin{figure}[bth]
\centering
\includegraphics[width=0.9\columnwidth]{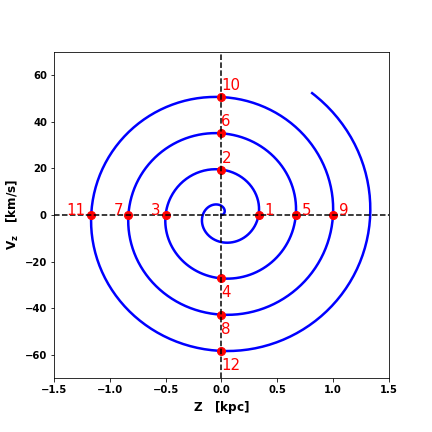}
\caption{
\label{fig:skecth}
Sketch plot of the intersections between the phase snail and $z$/$V_{z}$ axes. The blue line is an Archimedean spiral. The horizontal and vertical dashed lines represent the $z$ and $V_{z}$ axes, respectively. The indices of intersections are also labelled, with smaller indexes for inner intersections. The turning-around ($V_{z}=0$) and mid-plane points ($z=0$) are marked in odd and even numbers, respectively.}
\end{figure}

As shown in the sketch plot in Fig. \ref{fig:skecth}, the phase snail wraps up in the phase space and intersects with both $z$ and $V_{z}$ axes. The intersections on the $z$ axis, i.e. the odd points, reach their maximum vertical excursion with measurable $Z_{max}$. Similarly, the even intersections on the $V_{z}$ axis have measurable $V_{z,max}$, which corresponds to the vertical energy $E_{z}$. Besides, the intersections have increasing vertical energy $E_{z}$ and decreasing vertical frequency $\Omega_{z}$ from inner to outer regions of the phase space, with $Z_{max}$ and $V_{z,max}$ increasing monotonically. Thus, we can derive the $Z_{max}$ for an intersection on the $V_{z}$ axis by an interpolation method utilizing the two adjacent intersections on the $z$ axis with vertical action angle difference $\pm \pi/2$. For example, for point 2, its maximum vertical height can be determined by $Z_{max,2}= f(Z_{max,1},\ Z_{max,3})$, where $f$ is an arbitrary interpolation function. Similarly, we can derive the maximum vertical velocities for intersections on the $z$ axis with another interpolation function $g$, e.g. for point 3, $V_{z,max,3}= g(V_{z,max,2},\ V_{z,max,4})$.

%%%%%%%%%%%%%%%%%%%%%%%%%%%%%%%%%%%%%%%%%
%%%%%%%%%%%%%%%%%%%%%%%%%%%%%%%%%%%%%%%%%

Another important feature of the phase space motion is the linear relation between the vertical frequency $\Omega_{z}$ and the vertical action angle $\theta_{z}$:
%%%%%%%%%%%%%%%
\begin{equation}
\label{eq:tz_oz}
    \theta_{z}= \theta_{0} + \Omega_{z} \times {\rm T_{evol}} \ ,
\end{equation}
%%%%%%%%%%%%%%%
where $\theta_{0}$ and $\rm T_{evol}$ are the initial phase angle and the evolution time since the perturbation, respectively. The adjacent intersections on the $z$/$V_{z}$ axes have a constant action angle difference of $\Delta \theta_{z}= \pi/2$, which results from a constant vertical frequency difference $\Delta\,\Omega_{z}$ between the adjacent intersections. Thus, a parameter having a better linear relation with $\Omega_{z}$ would be preferred for the interpolation. For the intersections on the $V_{z}$ axis, both $V_{z,max}$ and $E_{z}$ can be used for the interpolation. As shown in the later Section \ref{sssec:oz_vz_zmax}, $V_{z,max}$ is better than $E_{z}$, and a linear interpolation function would be accurate enough and model independent.

After the ($Z_{max}$, $V_{z,max}$) pairs have been derived for the intersection points, the relation between $Z_{max}$ and $\Psi(Z_{max})= \frac{1}{2} V_{z,max}^{2}$ can be built at different radii, leading to a direct measurement of the Milky Way's vertical potential. Though we have utilized the same description of the phase space wrapping as \cite{Widmark2021a} (the potential measurement of which is embedded in the orbital integration as Eq. \ref{eq:Pz}), we do not make any assumption of the prior potential model as they do. Our measurement would be valuable to independently assess the popular Milky Way potential models, or to constrain the Milky Way global mass model.

%%%%%%%%%%%%%%%%%%%%%%%%%%%%%%%%%%%%%%%%%
\subsection{Method verification with test particle simulation}
\label{ssec:method_test}
%%%%%%%%%%%%%%%%%%%%%%%%%%%%%%%%%%%%%%%%%
To verify this model independent method, we run a test particle simulation similar to \cite{LiZY2021}. The initial condition sampling and 3-dimensional (3D) orbital integration are performed with AGAMA \citep{Vasiliev2019}. The input potential is the Model I from \cite{Irrgang2013}. We sample a thin disc tracer population utilizing the \textit{QuasiIsothermal} distribution function (DF) \citep{Binney2010, Binney2011}. The basic properties such as the radial and vertical scale lengths, velocity dispersion are set to represent a dynamically cold thin disk \citep{Binney2015, Bland-Hawthorn2016, LiZY2021}. Afterwards, the disc is perturbed following the impulse approximation by imposing a vertical downward velocity with the vertical position unchanged. The vertical velocity kick received by all the particles follows a Gaussian distribution of $-30\pm5\ {\rm km\,s^{-1} }$, i.e. the first approach used in \cite{LiZY2021}. The snapshot evolved for 500 Myr after the perturbation is adopted for the following analyses.

%%%%%%%%%%%%%%%%%%%%%%%%%%%%%%%%%%%%%%%%%
\subsubsection{Difference between $R-$ and $R_{g}$-binned snails}
\label{sssec:snail_diff}
\begin{figure*}[htbp]
\centering
\includegraphics[width=0.9\textwidth]{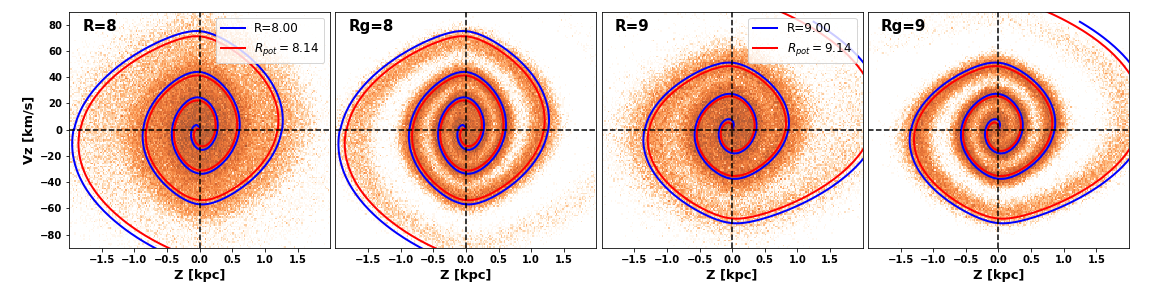}
\caption{
\label{fig:snail_diff}
The difference between $R$- and $R_{g}$-binned snails. The left two panels are for $R=8$ and $R_{g}= 8$ kpc, while the right two panels are for $R=9$ and $R_{g}= 9$ kpc, as labelled on the top left corners. The bin width is 0.6 kpc. All the maps are colored by the number density. The model snail shape curves from the phase-space orbit integration of the vertical potential grids (initially located at $V_{z}<0$ axis) are plotted by the solid lines. The blue lines are for the $R$-binned snails at the same radii. The red lines are from the potential radii corrected for the $R_{g}$-binned snails, with potential radii corrected from Fig. \ref{fig:Rg_Rpot}, which is about 0.14 kpc larger than the bin centers. Notice that, the blue lines are usually located in the outer edges of the $R_{g}$-binned snails, while the red lines are located close to the ridges of the snails. This verifies the necessity of potential radius correction for a $R_{g}$-binned snail.}
\end{figure*}

%%%%%%%%%%%%%%%%%%%%%%%%%%%%%%%%%%%%%%%%%
As illustrated in \cite{LiZY2021}, the vertical phase snails binned by the Galactocentric radius ($R$) and the guiding center radius ($R_{g}$) are different. The $R_{g}$-binned snails show higher clarity than the $R$-binned snails, showing more snail wraps, especially in the outer region of the phase space. This is because the $R_{g}$-binned stars have similar average orbital radii, and thus they feel similar vertical potential that leads to the winding of the phase snail. On the other hand, in the $R$-binned snail, stars have a wide guiding center radius distribution. The $R$-binned snail can be thought as a combination of different $R_{g}$-binned snails. Therefore, this $R_{g}$-mixing effect blurs the $R$-binned snail. This difference is clearly seen in Fig. \ref{fig:snail_diff}, in which $R$- and $R_{g}$-binned snails at two different radii from the test particle simulation are compared.

\cite{LiZY2021} found that the quantitatively measured shapes of $R$- and $R_{g}$-binned snails are similar. However, in order to derive the vertical potential with the snail intersections, we need to carefully quantify their difference. As an intuitive thought, stars of both snails should feel the same potential. In Fig. \ref{fig:snail_diff}, we overplot the toy model snails as the blue solid lines, which are built by the 1D vertical orbit integration under the vertical potential at $\Psi(z)|_{\rm R=8\, kpc}$ (left two panels) and $\Psi(z)|_{\rm R=9\, kpc}$ (right two panels), ignoring the radial and azimuthal motions. The model snails initially are energy grids and located at the negative $V_{z}$ axis, mimicking the impulse approximation with a negative velocity kick. For a star with a given vertical energy, we can transform it between $z-V_{z}$ space and $\Omega_{z}-\theta_{z}$ space, or conversely. We evolve the initial energy grids 500 Myr as the test particle simulation with the 1D vertical orbit integration, which then naturally wrap into the toy model snails. As shown in Fig. \ref{fig:snail_diff}, especially in the third panel, these toy model snails seem to follow the ridges of the R-binned snails well, especially in the inner region with lower vertical energies. On the other hand, in the very outer regions of the phase space, the toy model snails deviate from the snail density maps in the simulation, which is due to the ignoring of the in-plane motion. Nevertheless, this toy model snail derived from the 1D phase-space orbit integration works well in the inner region, where we can also clearly detect the phase snail in observations.

However, on the $R_{g}$-binned snail density maps, we find that these toy model snails, i.e. the blue lines in Fig. \ref{fig:snail_diff}, lie in the outer edges of the snails. It is more clearly illustrated in the inner radius of $R_{g}= 8$ kpc, i.e. the second panel in Fig. \ref{fig:snail_diff}. This confirms that the $R$- and $R_{g}$-binned snails are quantitatively different.

%%%%%%%%%%%%%%%%%%%%%%%%%%%%%%%%%%%%%%%%%
%%%%%%%%%%%%%%%%%%%%%%%%%%%%%%%%%%%%%%%%%

%%%%%%%%%%%%%%%%%%%%%%%%%%%%%%%%%%%%%%%%%
\subsubsection{Radius correction for the $R_{g}$-based phase snail}
\label{sssec:Rpot_simu}
%%%%%%%%%%%%%%%%%%%%%%%%%%%%%%%%%%%%%%%%%
Considering a pure circular orbit with a zero radial action $J_{R}= 0$, $R_{g}$ is exactly the Galactocentric radius R. For a star with a non-zero radial action, its orbit is rosette-like with radial oscillations around the guiding center radius. The star spends more time in the outer region of the guiding center radius because of the smaller velocity near the apocenter. Therefore, on average, the star feels a vertical potential at a radius slightly larger than $R_{g}$. We call this time weighted average orbital radius as the vertical potential radius $R_{pot}$.

In principle, this potential radius can be estimated under the Milky Way's potential for a star with a given radial action, which also varies for different stars. Thus, the calculation is model dependent and time consuming. Here, we propose an empirical approach to derive this potential radius based on the distribution of the guiding center radius of a stellar sample. Considering a narrow Galactocentric radial bin centered at $R$, the stars inside this bin should on average feel the vertical potential at $R$. The guiding center radius distribution of these stars would be skewed to a smaller radius. This is similar to the asymmetric drift effect. This Galactocentric radius $R$ can be considered as the potential radius $R_{pot}$ for these stars. Taking the medians of the $R_{g}$ distributions of stars at different radii, we can build an empirical relation between $R_{g}$ and $R_{pot}$. We then utilize an interpolation of this relation to relate a $R_{g}$-binned snail to a certain potential radius $R_{pot}$.

%%%%%%%%%%%%%%%%%%%%%%%%%%%%%%%%%%%%%%%%%
\begin{figure}[bth]
\centering
\includegraphics[width=0.9\columnwidth]{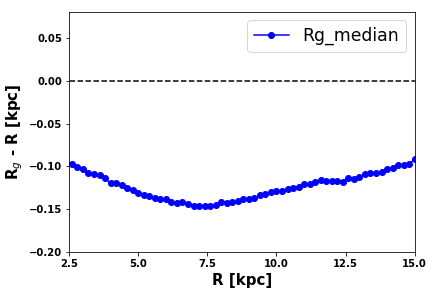}
\caption{
\label{fig:Rg_Rpot}
The difference between a radial bin center and its median guiding center radius ($R_{g}$) for the test particle simulation. The bin width of each radius is 0.6 kpc. This relation is utilized to relate a $R_{g}$-binned snail to the potential radius ($R_{pot}$).}
\end{figure}
%%%%%%%%%%%%%%%%%%%%%%%%%%%%%%%%%%%%%%%%%

In the test particle simulation, for particles located in different $R$ bins, the difference between the median of $R_{g}$ distribution and the Galactocentric radial bin center (i.e. $R_{pot}$) is shown in Fig. \ref{fig:Rg_Rpot}. The bin width is 0.6 kpc, which is the same in the following simulation analyses. Each Galactocentric radial bin usually corresponds to a $R_{g}$ distribution covering about 2-3 kpc. As a result, the median $R_{g}$ is about 0.1-0.15 kpc smaller than the bin center radius, i.e. the potential radius $R_{pot}$.

To verify this empirical correction, we overplot the toy model snails at $R_{pot}$ on Fig. \ref{fig:snail_diff} as the red lines. As shown in the second and fourth panels, the red lines are located closer to the ridge of the snail than the blue lines. They also stay in the inner regions of the blue lines, indicating a smaller or shallower vertical potential. Note that, in the very outer region, the model snails also deviate from the snail density maps, indicating that 1D orbital integration is not proper at such vertical height ($> 1.5$ kpc).

Ignoring this correction will result in a systematic underestimation in measuring the vertical potential from the $R_{g}$-binned phase snail, since the vertical potential profile at $R_{pot}$ is slightly shallower than that at $R_{g}$. This effect could be non-negligible at smaller radii, where the potential changes significantly with a small radius difference. In addition, this test particle simulation imitates a thin disc tracer population. For the observations, the difference between the potential radius $R_{pot}$ and $R_{g}$ could be larger. The correction also depends on the sample selection, such as selecting stars with colder orbits will lead to a smaller difference between $R_{pot}$ and $R_{g}$.

%%%%%%%%%%%%%%%%%%%%%%%%%%%%%%%%%%%%%%%%%
\subsubsection{Refined linear interpolation method for snail intersections}
\label{sssec:oz_vz_zmax}
\begin{figure*}[htbp]
\centering
\includegraphics[width=1.0\textwidth]{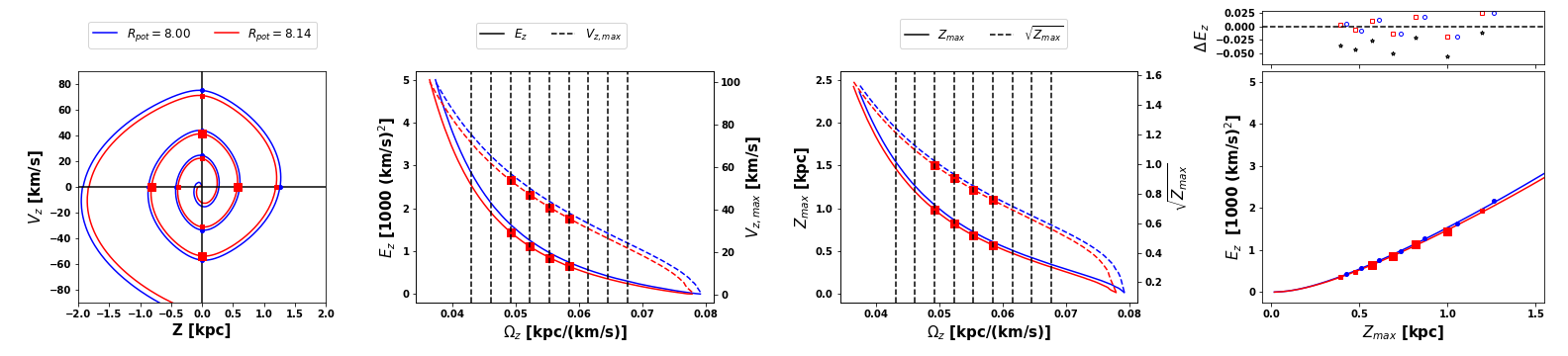}
\caption{
\label{fig:oz_vz_zmax}
Vertical potential derived from idealized phase snail calculated from the 1D orbital integration. The blue and red lines represent the snail shape curves at $R_{pot}= 8$ kpc and $R_{pot}= 8.14$ kpc, corresponding to the snails binned by $|R-8|< 0.3$ kpc and $|R_{g}-8|< 0.3$ kpc, respectively. \textbf{Left panel}: snail shape curves in the $z-V_{z}$ phase space, with snail intersections marked by dots (for $z < 1.5$ kpc). Four intersections labelled by the larger red squares are marked in all panels to help understand the interpolation method. \textbf{Middle left panel}: the vertical energy ($E_{z}$ in the left axis, solid lines) and the maximum vertical velocity ($V_{z,max}$ in the right axis, dashed lines) as a function of the vertical frequency ($\Omega_{z}$) for the two snails shown in the left panel. The vertical dashed lines indicate the frequencies of the intersections. \textbf{Middle right panel}: same as the middle left panel but for the maximum vertical height ($Z_{max}$ in the left axis, solid lines) and its square root ($\sqrt{Z_{max}}$ in the right axis, dashed lines). Note that within the region covered by the vertical lines, the dashed lines show better linear relations with the vertical frequency than the solid lines. \textbf{Right panel}: snails presented as $E_{z}$ versus $Z_{max}$, i.e. the vertical potential profiles. The filled symbols are derived utilizing a linear interpolation function based on the relations between $\Omega_{z}$ and $V_{z,max}$ and $\sqrt{Z_{max}}$ (dashed lines in the middle panels), for the intersections marked in the left panel. In the top panel, the open symbols are the relative residuals between the filled symbols and the model potential curves. The black symbols are the relative residuals between the red filled symbols and the blue solid line, which shows a systematical underestimation for using a $R_{g}$ binned snail without considering the potential radius correction.}
\end{figure*}

%%%%%%%%%%%%%%%%%%%%%%%%%%%%%%%%%%%%%%%%%
As illustrated in Fig. \ref{fig:skecth}, our method relies on the correct interpolations of the $Z_{max}$ and $V_{z,max}$ values for the intersections between the phase snail and the $V_{z}$ axis and $z$ axis, respectively. A straightforward approach for the interpolation is to directly use the $Z_{max}$ or $V_{z,max}$ values of the adjacent intersections. For example, $Z_{max}$ of point 4 can be estimated with the average $Z_{max}$ of points 3 and 5, and $V_{z,max}$ of point 3 can be estimated with linear interpolation of $V_{z,max}$ of points 2 and 4. However, in order to derive the vertical potential with less systematical errors, physical quantities linearly related to the vertical frequency $\Omega_{z}$ are better choices for the interpolation. We test different parameters using the model snails with the results shown in Fig. \ref{fig:oz_vz_zmax}. The model snails at $R_{pot}= 8$ kpc and $R_{pot}= 8.14$ kpc (corresponding to the snail at $R_{g}= 8$ kpc) are displayed as the blue and red solid lines in the left panel of Fig. \ref{fig:oz_vz_zmax}, respectively. The intersections are defined by the action angles where $\theta_{z}= \frac{k}{2}\pi$ ($k$ is an integer). They are marked as dots in the left panel, and their corresponding $\Omega_{z}$ are indicated by the vertical dashed lines in the middle panels of Fig. \ref{fig:oz_vz_zmax}.

%%%%%%%%%%%%%%%%%%%%%%%%%%%%%%%%%%%%%%%%%
%%%%%%%%%%%%%%%%%%%%%%%%%%%%%%%%%%%%%%%%%

For the $z$-axis intersections, we could utilize $V_{z,max}$ or $E_{z}= \frac{1}{2}\, V_{z,max}^{2}$ as the linear interpolation quantity. The relations of $\Omega_{z}$ with $E_{z}$ and $V_{z,max}$ are shown as the solid and dashed lines in the second panel of Fig. \ref{fig:oz_vz_zmax}, respectively. Apparently, $E_{z}$ has a non-linear dependence on $\Omega_{z}$, while $V_{z,max}$ has a roughly good linear relation with $\Omega_{z}$, in the region considered. Therefore, using $E_{z}$ as the linear interpolation quantity will result in an overestimation, while utilizing $V_{z,max}$ can significantly reduce the systematic error in the interpolation. For this reason, we adopt $V_{z,max}$ as the linear interpolation quantity to derive $\Psi(Z_{max})= E_{z}$ for the $z$-axis intersections.

For the $V_{z}$-axis intersections, we explore the relation between $Z_{max}$ and $\Omega_{z}$, shown as the solid lines in the third panel of Fig. \ref{fig:oz_vz_zmax}. It is also a curved relation, similar to the relation between $E_{z}$ and $\Omega_{z}$. Thus, using $Z_{max}$ directly as the linear interpolation quantity will result in an overestimation. As there is squared relation between $E_{z}$ and $V_{z,max}$, we explore the relation between the square root of $Z_{max}$, i.e. $\sqrt{Z_{max}}$, and $\Omega_{z}$, illustrated by the dashed lines in the third panel of Fig. \ref{fig:oz_vz_zmax}. The relation between $\sqrt{Z_{max}}$ and $\Omega_{z}$ seems linear in the region covered by the vertical dashed lines. Thus, $\sqrt{Z_{max}}$ is adopt as the linear interpolation quantity to derive $Z_{max}$ for the $V_{z}$-axis intersections.

Though the above two approximate linear relations are derived under the potential of \cite{Irrgang2013}, they should remain valid in real observations for two reasons. First, the shapes of popular Milky Way potentials are similar after shifting to the same Solar radius. Second, in other radial bins, there are also good linear relations of $\Omega_{z}$ with $\sqrt{Z_{max}}$ and $V_{z,max}$. The linear interpolation of these two quantities should work well at different radii. As a summary, we refine our interpolation method for the $(i+1)$-th intersections as following:
%%%%%%%%%%%%%%%%%%%%%%%%%%%%%%%%%%%%%%%%%
\begin{equation} 
\label{eq:interp} 
\left\{
\begin{aligned} 
   \sqrt{Z_{max,i+1}} & = \frac{1}{2}( \sqrt{Z_{max,i}} + \sqrt{Z_{max,i+2}} ) \quad & if\quad z_{i+1}=0 \, ;\\
   V_{z,max,i+1} & = \frac{1}{2}( V_{z,max,i} + V_{z,max,i+2} ) \quad & if\quad V_{z,i+1}=0 \, .
\end{aligned} 
\right. 
\end{equation}
%%%%%%%%%%%%%%%%%%%%%%%%%%%%%%%%%%%%%%%%%
$Z_{max,i}$ and $V_{z,max,i}$ are the absolute values measured for the intersections with phase space coordinates of ($z_{i},\, V_{z,i}$) on the $z$-axis and $V_{z}$-axis, respectively.  

Combining with Eq. \ref{eq:Ez_zvzmax}, we can obtain the ($Z_{max}$, $\Psi(Z_{max})$) for $N$-2 intersections (excluding the first and last snail intersections), where $N$ is the number of intersections detected for a snail. The interpolation is performed for the intersections measured from the model snails, and the resultant potential data points are shown as the filled symbols in the right panel of Fig. \ref{fig:oz_vz_zmax}. They have a good consistency with the input potential profiles, with residuals quite close to zero, as illustrated by the open symbols. However, ignoring the difference between $R_{g}$ and $R_{pot}$ will result in a systematic underestimation of up to 5\%, as indicated by the black open symbols in the top right panel of Fig. \ref{fig:oz_vz_zmax}. 

According to the 1D model snail test, the interpolation method defined by Eq. \ref{eq:interp} is accurate, showing much less systematic uncertainty. In real observation, the main error source in the observation should arise from the determination of these intersection positions, and the ignorance of the difference between the vertical potential profiles at $R_{g}$ and $R_{pot}$. The derived ($Z_{max}$, $\Psi(Z_{max})$) potential points provide a direct and model independent measurement of the vertical potential profile of our Milky Way. It can be compared with the popular Milky Way potentials, and applied with a further mass modelling to construct the Galactic mass distribution.

%%%%%%%%%%%%%%%%%%%%%%%%%%%%%%%%%%%%%%%%%
\subsubsection{Intersections measurement and uncertainties}
\label{sssec:simu_inter_phiz}
%%%%%%%%%%%%%%%%%%%%%%%%%%%%%%%%%%%%%%%%%
This section describes our method to measure the snail intersections with both axes. We start with the test particle simulation. Although the $R$-binned snails on average feel the vertical potential requiring no potential radius correction, they are more dispersed than the $R_{g}$-binned snails. In order to reduce the errors of intersection measurements, we prefer to measure the intersections based on the $R_{g}$-binned snails and apply the interpolation method, which will also be performed on the observational data in the next section.

\begin{figure*}[htbp]
\centering
\includegraphics[width=0.9\textwidth]{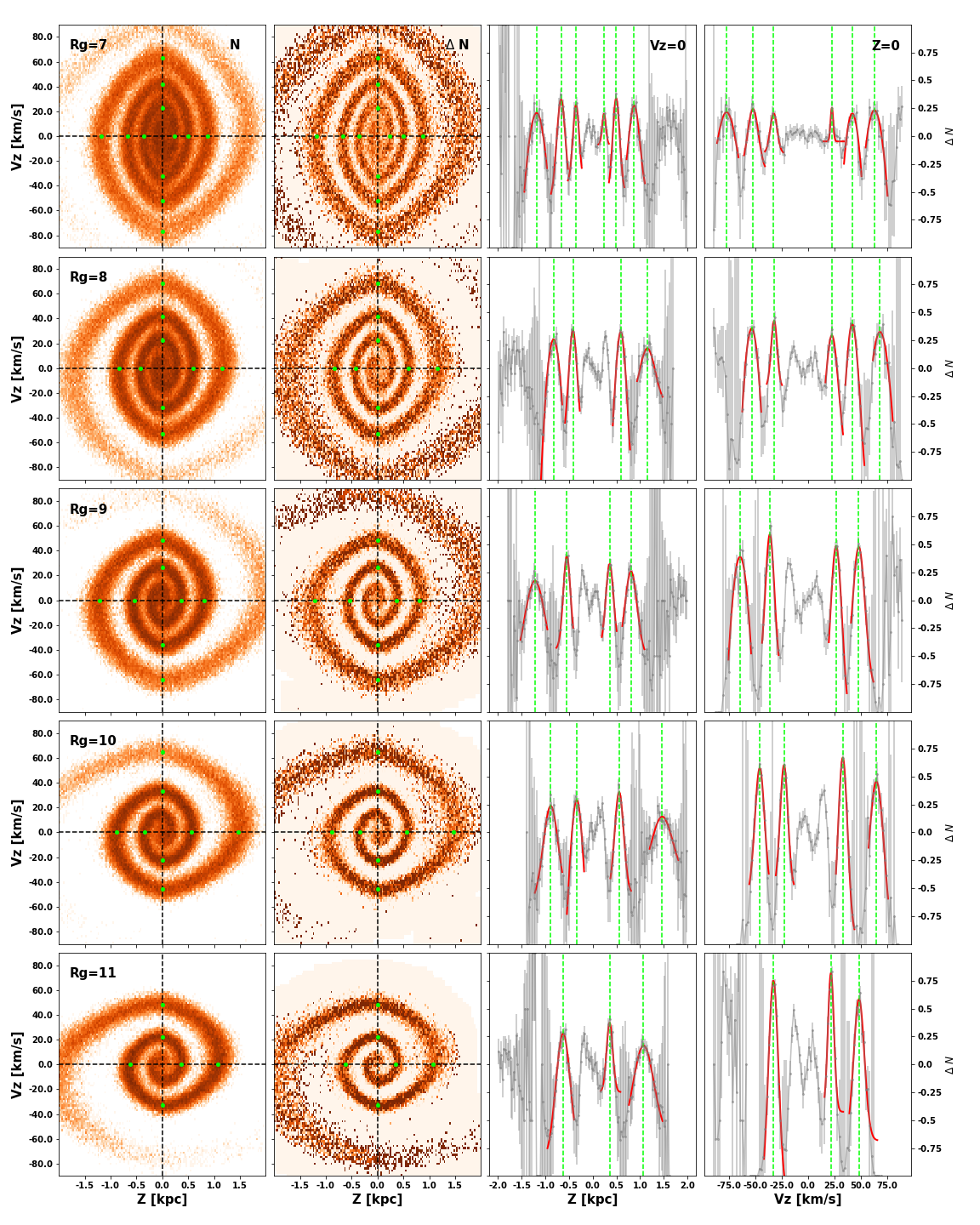}
\caption{
\label{fig:simu_inter}
Measure the intersections for the simulated phase snails. The left column shows the snails at different guiding center radii (as indicated), while the second column presents their density contrast maps. The pixel size is 0.026 kpc $\times$ 1.8 $\rm km\,s^{-1}$. The third and fourth columns display the density contrast profiles (grey lines) for stripes at $V_{z}=0$ (along $z$ axis) and $z=0$ (along $V_{z}$ axis), with widths of 20 $\rm km\,s^{-1}$ and 0.3 kpc, respectively. The error bars show the numerical noise. The red lines show the Gaussian profiles fitting to the visually selected intersection regions, with the Gaussian peaks indicated by the light green dashed lines. The intersections, i.e. the Gaussian peaks are marked in the left two columns with the light green dots.}
\end{figure*}

We take five radial bins centered at $R_{g}= 7, 8, 9, 10, 11$ kpc with the bin width of 0.6 kpc. The phase space density distributions of the snails are shown in the left column of Fig. \ref{fig:simu_inter}, with a pixel size of 0.026 kpc $\times$ 1.8 $\rm km\,s^{-1}$. In observations, the snail number density map is usually convolved with a Gaussian kernel to get a smooth background \citep{Laporte2019, LiZY2020}. The density contrast map $\Delta N$ after subtracting the smooth background significantly enhances the snail signal. We also show the density contrast maps (with a Gaussian kernel dispersion of $\sigma= 3$ pixels) for our $R_{g}$-binned snails in the second column of Fig. \ref{fig:simu_inter}. 

In order to measure the intersections, we extract the number density profiles along two stripes at $V_{z}=0$ (along $z$ axis) and $z=0$ (along $V_{z}$ axis), with widths of 20 $\rm km\,s^{-1}$ and 0.3 kpc, respectively. These widths are similar to the widths of the snails in the $z$ and $V_{z}$ directions. Similar to the 2D density contrast map, we perform 1D Gaussian kernel convolution to the extracted number density profiles. The density contrast profiles are displayed in the third and fourth columns of Fig. \ref{fig:simu_inter}, for the two stripes at $V_{z}=0$ and $z=0$, respectively. In these 1D number density contrast profiles, there are several peaks related to the locations of intersections. By visual inspection, we select the rough regions where peaks emerge. We ignore several peaks located in the inner region of the phase space, where the peaks could be blurred by the high density background in observations. For these regions, we fit the density contrast profiles with a Gaussian function with the peak corresponding to the location of an intersection. The best-fit Gaussian profiles are shown as the red lines in the right two columns of Fig. \ref{fig:simu_inter}, while the intersections are indicated by the light green dashed lines. The error bars of the intersections are taken from the maximum of the error from the Gaussian fitting and the pixel size, of which the former is usually smaller than the latter. The measured intersections are also marked as the light green dots in the original and contrast density maps, shown in the left two columns of Fig. \ref{fig:simu_inter}.

%%%%%%%%%%%%%%%%%%%%%%%%%%%%%%%%%%%%%%%%%
%%%%%%%%%%%%%%%%%%%%%%%%%%%%%%%%%%%%%%%%%

\begin{figure*}[bth]
\centering
\includegraphics[width=0.9\textwidth]{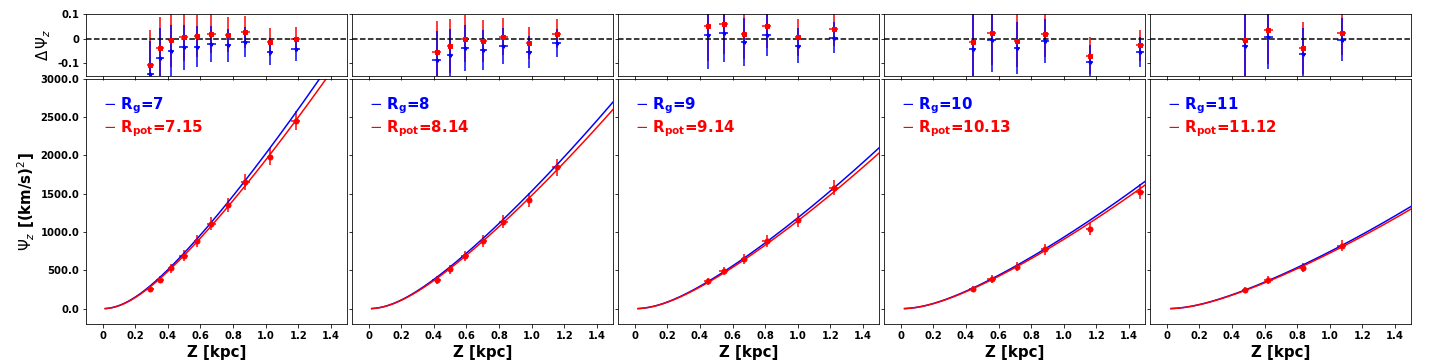}
\caption{
\label{fig:simu_phiz}
Vertical potential derived from the interpolation of the phase snail intersections measured from the test particle simulation. Lower panels: The blue lines show the model potential measured at $R= R_{g}$, while the red lines display the potential at $R= R_{pot}$ that is corrected from the guiding center radius with the relation shown in Fig. \ref{fig:Rg_Rpot}. The red dots are derived from the interpolation of the intersections between the snails and the $z/V_{z}$ axes. Upper panels: The relative residuals between the potentials derived from the snail intersections and the model potentials at radii of $R= R_{g}$ (blue stars) and $R= R_{pot}$ (red squares).}
\end{figure*}

For the measured intersections, we can apply the interpolation method in Eq. \ref{eq:interp} to obtain vertical potential measurements at several vertical heights. The results are shown in Fig. \ref{fig:simu_phiz}. In the lower panels, the interpolated points are well consistent with the vertical potentials at the corresponding potential radii, i.e. the red lines, which are derived from the relation displayed in Fig. \ref{fig:Rg_Rpot}. As explained previously, these points show systematic underestimation, compared to the potential just at the guiding center radii (the blue lines). This underestimation is more significant at smaller radii. The residuals between the derived potential data points and the potential profiles at $R_{pot}$ and $R_{g}$ are indicated by the red and blue points in the top panels of Fig. \ref{fig:simu_phiz}, respectively. Compared to the potential at $R_{pot}$, the interpolated potential points are consistent with deviations less than 5\%. However, compared to the vertical potential profiles at $R_{g}$, they manifest systematic underestimation with slightly shallower profiles, up to $\sim 5$\% at the inner most radius. Note that, our test particle simulation mimics a cold thin disc, which shows only about 0.1-0.15 kpc difference between $R_{g}$ and $R_{pot}$. In the observations, this potential radius correction could be larger, resulting in a more significant underestimation. As shown by the error bars in the upper panels of Fig. \ref{fig:simu_phiz}, the uncertainty due to the intersections measurement is about 10\%. As we do not include background particles in the measurement, this uncertainty could also be larger in the observations.

%%%%%%%%%%%%%%%%%%%%%%%%%%%%%%%%%%%%%%%%%
%%%%%%%%%%%%%%%%%%%%%%%%%%%%%%%%%%%%%%%%%

%%%%%%%%%%%%%%%%%%%%%%%%%%%%%%%%%%%%%%%%%%%%%%%%%%%%%%%%%%%%%%
\section{Data} 
\label{sec:data}
%%%%%%%%%%%%%%%%%%%%%%%%%%%%%%%%%%%%%%%%%%%%%%%%%%%%%%%%%%%%%%
We try to obtain the disk vertical potential measurements with the interpolation method of the phase snail intersections based on Gaia Data Release 3 \citep{GaiaDR32022}. Gaia DR3 RVS provides radial velocity measurements for about 33.8 million objects. We take the geometric distances from \cite{Bailer-Jones2021}, which are based on a Bayesian approach that uses a prior constructed from a three-dimensional model of the Milky Way. We apply a distance quality cut, requiring the relative parallax error ($\sigma_{\varpi}$) smaller than 20 per cent (i.e. $\varpi/\sigma_{\varpi}>5$). In terms of the astrometric measurements, we require the renormalised unit weight error (RUWE) to be smaller than 1.4. The final sample contains 27,209,191 stars.

%%%%%%%% Note to revise it according the sample used
The coordinates system adopted in this paper is the classical Galactocentric cylindrical system ($R,\, \phi,\, z$), with the Galactocentric radius $R$ increasing radially outward, $\phi$ increasing in the direction of Galactic rotation, and positive $z$ pointing toward the North Galactic Pole. The Sun is located at ($-8.34, 0, 0.027$) kpc in the Galactic Cartesian coordinates system \citep{ChenB1999, Reid2014b}. The Solar motion relative to the local standard of rest (LSR) is $\rm (U_{\odot}, V_{\odot}, W_{\odot}) = (7.01, 12.09, 7.01)\ km\, s^{-1}$ \citep{HuangY2016}, and the circular velocity of LSR is adopted as 240 km s$^{-1}$ \citep{Reid2014b}.

%%%%%%%%%%%%%%%%%%%%%%%%%%%%%%%%%%%%%%%%%%%%%%%%%%%%%%%%%%%%%%
\section{Results} 
\label{sec:results}
%%%%%%%%%%%%%%%%%%%%%%%%%%%%%%%%%%%%%%%%%%%%%%%%%%%%%%%%%%%%%%
%%%%%%%%%%%%%%%%%%%%%%%%%%%%%%%%%%%%%%%%%
\subsection{Measuring the snail intersections in the observation}
\label{ssec:obs_inter}
%%%%%%%%%%%%%%%%%%%%%%%%%%%%%%%%%%%%%%%%%
Although $R$-binned snails directly represent the vertical potential, they are more dispersed than $R_{g}$-binned snails. In order to reduce the uncertainty in measuring the snail intersections, we utilize $R_{g}$-binned snails in observations. The guiding center radius $R_{g}$ is calculated as the radius of a circular orbit with the same angular momentum ($L_{z}$) as the star, i.e. $R_{g}=RV_{\phi} \big/ V_{c}(R_{g})$, where $V_{\phi}$ is the azimuthal velocity and $V_{c}$ is the rotation curve. We utilize the rotation curve from \cite{HuangY2016}, with a circular velocity of $240 \pm 6$ km s$^{-1}$ at the Solar radius ($R_{0}$=8.34 kpc), to build a relation of $L_{z}-R_{g}$. By mapping the observed angular momentum of a star to the derived $L_{z} - R_{g}$ relation, we can obtain its $R_{g}$ through an interpolation. We explore nine radial bins with centers at $R_{g}= 7.5, 8., 8.34, 8.5, 9, 9.5, 10, 10.5, 11$ kpc, with a bin width of $\Delta R_{g}= 0.5$ kpc. More inner radii are not explored in this work as the snails could be more complicated, while at the outer radii, measurements could be dominated by the noise. We trace the snails in a phase space region of $|z|$< 1.5 kpc and $|V_{z}|< 75\ {\rm km\,s^{-1}}$, as the outer region is dominated by the numerical noise.

\begin{figure*}[htbp]
\centering
\includegraphics[width=0.8\textwidth]{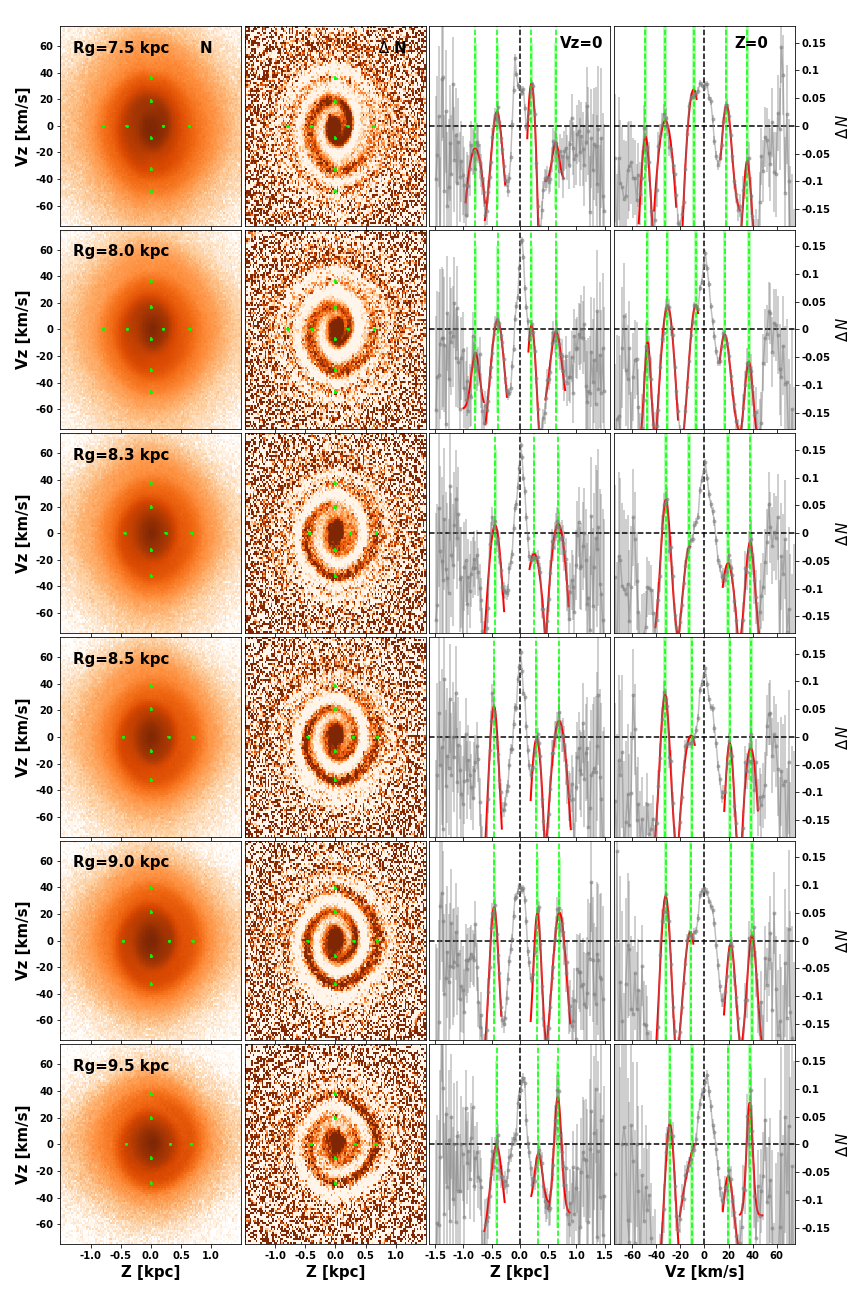}
\caption{
\label{fig:obs_inter}
Same as Fig. \ref{fig:simu_inter} but for the measurement of the intersections for the observed phase snails. The pixel size is 0.025 kpc $\times$ 1.5 $\rm km\,s^{-1}$.}
\end{figure*}

The phase space density distributions of $R_{g}$-binned snails are shown in the left column of Fig. \ref{fig:obs_inter}, with their density contrast maps displayed in the second column. Note that, compared to the simulated snails, the observed ones contain large fractions of background stars. Nevertheless, we have one order of magnitude larger sample size. The left two columns show clear radial variation of the phase snails, as snails at outer radii are more elongated along $z$ axis. This is due to the fact that the outer radius has shallower vertical potential, which results in a larger $Z_{max}$ for a given vertical energy. The density contrast profiles along $z$ and $V_{z}$ axes are shown in the right two columns of Fig. \ref{fig:obs_inter}. The snail intersections are measured in the same way as in the test particle simulation, and are illustrated by the vertical dashed lines in the right two columns and the light green points in the second column of Fig. \ref{fig:obs_inter}. 

%%%%%%%%%%%%%%%%%%%%%%%%%%%%%%%%%%%%%%%%%
%\begin{figure*}[htbp]
%\centering
%\includegraphics[width=0.8\textwidth]{Rg_snail_zVz_contrast_profiles.png}
%\caption{
%\label{fig:obs_inter}
%Measure the intersections for the observed phase snails, same as the Fig. \ref{fig:simu_inter}.}
%\end{figure*}
%%%%%%%%%%%%%%%%%%%%%%%%%%%%%%%%%%%%%%%%%
%%%%%%%%%%%%%%%%%%%%%%%%%%%%%%%%%%%%%%%%%
%%%%%%%%%%%%%%%%%%%%%%%%%%%%%%%%%%%%%%%%%
\renewcommand{\thefigure}{\arabic{figure} (Continued.)}
\addtocounter{figure}{-1}
%%%%%%%%%%%%%%%%%%%%%%%%%%%%%%%%%%%%%%%%%
\begin{figure*}[htbp]
\centering
\includegraphics[width=0.8\textwidth]{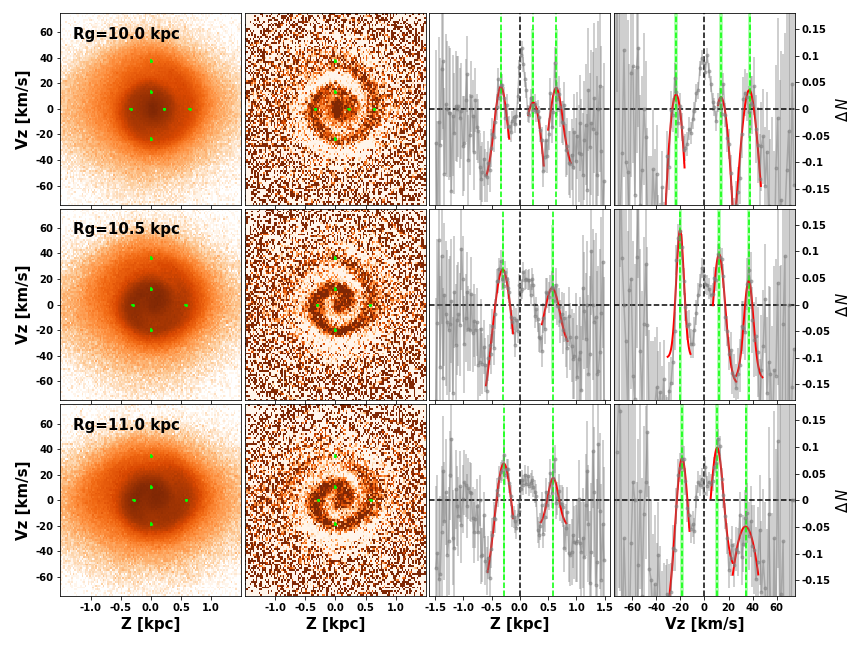}
\caption{}
\end{figure*}
%%%%%%%%%%%%%%%%%%%%%%%%%%%%%%%%%%%%%%%%%
\renewcommand{\thefigure}{\arabic{figure}}
%%%%%%%%%%%%%%%%%%%%%%%%%%%%%%%%%%%%%%%%%

%%%%%%%%%%%%%%%%%%%%%%%%%%%%%%%%%%%%%%%%%
\begin{figure}[bth]
\centering
\includegraphics[width=0.9\columnwidth]{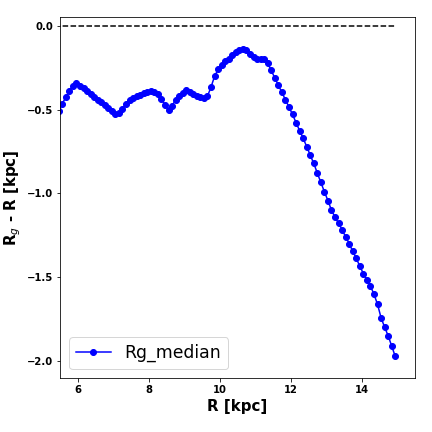}
\caption{
\label{fig:obs_Rpot}
The same as Fig. \ref{fig:Rg_Rpot} but for the observational data. The bin width of each point is 0.5 kpc. The guiding center radius is calculated according to the rotation curve from \cite{HuangY2016}.}
\end{figure}
%%%%%%%%%%%%%%%%%%%%%%%%%%%%%%%%%%%%%%%%%

The empirical relation between the guiding center radius and the potential radius for the observational sample is shown in Fig. \ref{fig:obs_Rpot}. The corresponding angular momenta ($L_{z}$) and the potential radii ($R_{pot}$) for the $R_{g}$ bins used are listed in Table \ref{tab:RgLzRpot}. The bin width of each $R$ bin is 0.5 kpc, which is the same as the $R_{g}$ bin width. In the region of $6<R<10$ kpc, the guiding center radius is about $0.4-0.5$ kpc smaller than the potential radius. This difference is significantly larger than that in the test particle simulation, which could thus result in a significant underestimation if ignoring this $R_{pot}$ correction. At $R\sim 11$ kpc, the difference reaches a minimum of about 0.2 kpc. In the outer ($R>12$ kpc) region, the relation shows a rapid decreasing pattern, which could be due to insufficient stars. The relation in the inner region is not shown due to the truncation of the rotation curve of \cite{HuangY2016} at about 5.5 kpc. Note that, this $R_{g}\ vs.\ R_{pot}$ relation is empirical, and should change with the sample selection. For example, if we choose stars in colder orbits, the difference between $R_{g}$ and $R_{pot}$ would be smaller. However, such choice would reduce the number statistics to allow for the accurate snail shape measurement.

%%%%%%%%%%%%%%%%%%%%%%%%%%%%%%%%%%%%%%%%%
%\begin{landscape}
\begin{table*}
\renewcommand\arraystretch{1.3}
\centering
\caption{The corresponding angular momenta ($L_{z}$) and the potential radii ($R_{pot}$) for the $R_{g}$ bins used in our sample.}
%\small
\label{tab:RgLzRpot}
\begin{tabular}{ c | c c c c c c c c c}
\hline
\hline
$R_{g}\ [{\rm kpc}]$   &  7.5  &  8.0  &  8.34  &  8.5  &  9.0  &  9.5  &  10.0 
        &  10.5 & 11.0  \\
\hline
  $L_{z}\ [{\rm kpc\,km\,s^{-1}}]$  & 1821.2  &  1945.5  &  2000.7  & 2027.3  & 2123.9  &  2205.7  &  2288.9  &  2361.6  &  2473.1  \\
 $R_{pot}\ [{\rm kpc}]$  &  7.90  &  8.48  &  8.78  &  8.91  &  9.42  &  9.82  &
               10.20  & 10.64  &  11.20  \\
 \hline
\end{tabular}
\end{table*}
%\end{landscape}
%%%%%%%%%%%%%%%%%%%%%%%%%%%%%%%%%%%%%%%%%

Based on the measured snail intersections from the observation, we apply the interpolation method as defined by Eq. \ref{eq:interp} to obtain $Z_{max}$ and $\Psi(Z_{max})= \frac{1}{2}V_{z,max}^{2}$ of these intersections. The derived snail potential data points are shown in Fig. \ref{fig:comp_phiz} as the red dots. These data points provide a model independent assessment of Milky Way's vertical potential, and can also be applied to the further mass modelling.

%%%%%%%%%%%%%%%%%%%%%%%%%%%%%%%%%%%%%%%%%
\subsection{Comparison with the popular Milky Way potentials}
\label{ssec:obs_comp}
\begin{figure*}[htbp]
\centering
\includegraphics[width=0.9\textwidth]{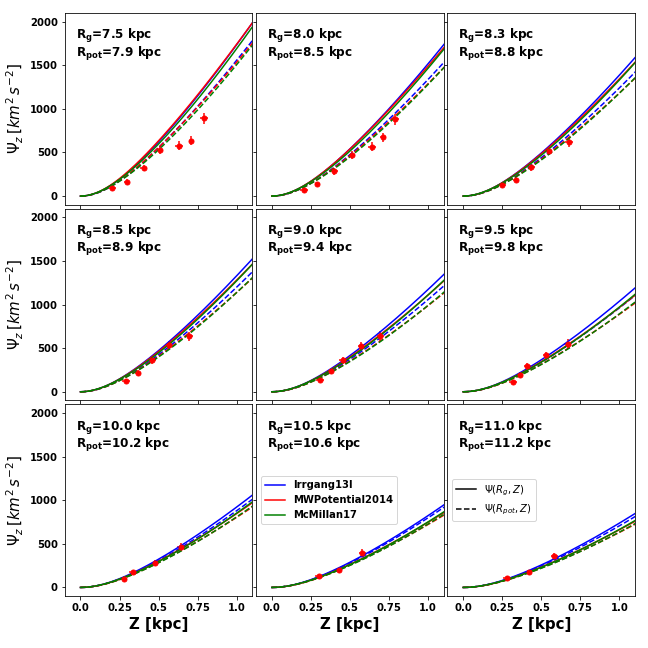}
\caption{
\label{fig:comp_phiz}
Comparison between the obtained vertical potential of the snail intersections and the popular Milky Way potentials. The red points are the interpolated snail intersections utilizing Eq. \ref{eq:interp}. The colored lines are the vertical potential profiles from three popular Milky Way potentials, i.e. blue for the Model I in \cite{Irrgang2013}, red for the potential from \cite{Bovy2015} (MWPotential2014 in galpy), and green for the potential from \cite{McMillan2017}. These vertical potential profiles have been radially shifted according to the different Solar radii ($R_{0}$) used. The solid lines show the potential at each $R_{g}$ bin center (as labelled), while the dashed lines display the potential at the corresponding potential radii $R_{pot}$ (as labelled), corrected from the $R_{g}$ values according to Fig. \ref{fig:obs_Rpot}.}
\end{figure*}

%%%%%%%%%%%%%%%%%%%%%%%%%%%%%%%%%%%%%%%%%
We compare our measurements of the Milky Way potential with three popular models of it, which are the Model I from \cite{Irrgang2013} (Irrgang13I), MWPotential2014 from \cite{Bovy2015}, and McMillan17 from \cite{McMillan2017}, shown as the blue, red and green lines, respectively, in Fig. \ref{fig:comp_phiz}. As these potentials have different Solar positions, they are radially shifted to the same Solar radius, i.e. $R_{0}= 8.34$ kpc. Besides the vertical potentials at the centers of $R_{g}$ bins (solid lines), the potentials at the corresponding potential radii are also displayed with the dashed lines in Fig. \ref{fig:comp_phiz}. The potential radii are calculated from the empirical relation shown in Fig. \ref{fig:obs_Rpot}.

The potential points derived from the snail intersections agree very well with the other popular Milky Way potentials from 7.5 to 11 kpc at $|z| < 1$ kpc. The interpolated snail intersections are slightly shallower than the vertical potential profiles at $R_{g}$, and are more consistent with those profiles at the corresponding potential radii. This phenomenon is more obvious at inner radii, where the vertical potential difference between $R_{g}$ and $R_{pot}$ is larger. At outer radii, as the radial gradient of the potential becomes shallower, the vertical potential profiles at $R_{g}$ and $R_{pot}$ are similar to each other. For the current snail intersection measurements, we only explore low vertical regions, which are only $|z| \lesssim 1$ kpc as shown in the second column of Fig. \ref{fig:obs_inter}. In addition, we have no interpolation for the outermost intersections in the phase space, which would result in smaller regions of the interpolated snail potential points.

Note that, at the inner two radii, i.e. $R_{g}= 7.5, 8.0$ kpc, there are two data points at $z \sim 0.7$ kpc showing significant deviations from the Milky Way potential profiles. This is likely due to the $Z_{max}$ estimation uncertainties for the intersections at $z \sim 0.7$ kpc. As shown in the second panel of the top row in Fig. \ref{fig:obs_inter}, the snail density contrast map at $z \sim 0.7$ kpc seems to show two adjacent wraps, with the outer wrap deviating from the snail pattern of the inner wrap. \cite{Antoja2022} also found an additional wrap at $V_{z} \sim 50\ {\rm km\,s^{-1}}$ in the volume at $R=8$ kpc, while our additional snail wrap is more clearly seen at $z \sim 0.7$ kpc and $V_{z} \sim 0\ {\rm km\,s^{-1}}$. The outer wrap is more obvious, which results in a larger $Z_{max}$ for that snail intersection. The interpolation transfers this overestimation to the two adjacent snail intersections on $V_{z}$ axis. If we take this $Z_{max}$ overestimation into consideration, the relevant interpolated snail intersections will become closer to the dashed lines in Fig. \ref{fig:comp_phiz}.

%%%%%%%%%%%%%%%%%%%%%%%%%%%%%%%%%%%%%%%%%
%%%%%%%%%%%%%%%%%%%%%%%%%%%%%%%%%%%%%%%%%

%%%%%%%%%%%%%%%%%%%%%%%%%%%%%%%%%%%%%%%%%
\subsection{Modelling the global disc}
\label{ssec:mod_global}
%%%%%%%%%%%%%%%%%%%%%%%%%%%%%%%%%%%%%%%%%
Based on the direct measurements of the vertical potential from the interpolated snail intersections, we perform mass modelling to these potential data points, to constrain the mass distributions of different Galactic components. We apply a mass model consisting of double exponential discs and an NFW \citep{NFW1996} dark matter halo. For simplicity, we regard the two discs as very flattened systems, and integrate the density profile along the $z$ direction twice to directly obtain their reduced vertical potential. See the detailed derivation in Appendix \ref{sec:app_mglobal}. The total reduced vertical potential at ($R$, $z$) is:
%%%%%%%%%%%%%%%
\begin{equation}
\label{eq:gol_dexp}
\begin{aligned}
\Psi(R, z) &=\frac{GM_{d1}}{R_{d1}^2} \exp \left(-\frac{R}{R_{d1}}\right)\left[z-z_{d1}+z_{d1} \exp \left(-\frac{z}{z_{d1}}\right)\right] \\
&+\frac{GM_{d2}}{R_{d2}^2} \exp \left(-\frac{R}{R_{d2}}\right)\left[z-z_{d2}+z_{d2} \exp \left(-\frac{z}{z_{d2}}\right)\right] \\
&+GM_h\left[\frac{\ln \left(1+\frac{R}{a_h}\right)}{R}-\frac{\ln \left(1+\frac{r}{a_h}\right)}{r}\right] \ ,
\end{aligned}
\end{equation}
%%%%%%%%%%%%%%%
where $r= \sqrt{R^{2}+z^{2}}$ and $G$ is the gravitational constant. $M_{d1}$, $R_{d1}$ and $z_{d1}$ are the total mass, the scale length and scale height of the thin disc, respectively, while $M_{d2}$, $R_{d2}$ and $z_{d2}$ are for the thick disc. $M_{h}$ and $a_{h}$ are the scale mass and scale radius for the dark matter halo. We use the Markov Chain Monte Carlo (MCMC) technique to fit the interpolated snail intersections and obtain the parameters' estimation. The MCMC package we use is {\tt EMCEE} \citep{Foreman-Mackey2013}. As our snail potential data points are at low vertical regions, we fix the scale heights of discs as $z_{d1}= 0.3$ kpc and $z_{d2}= 0.9$ kpc. We also apply Gaussian priors on the radial scale lengths, which are $2.6 \pm 0.52$ kpc and $3.6 \pm 0.72$ kpc for the thin and thick discs, respectively \citep{Bland-Hawthorn2016, McMillan2017}. 

\begin{figure*}[htbp]
\centering
\includegraphics[width=0.9\textwidth]{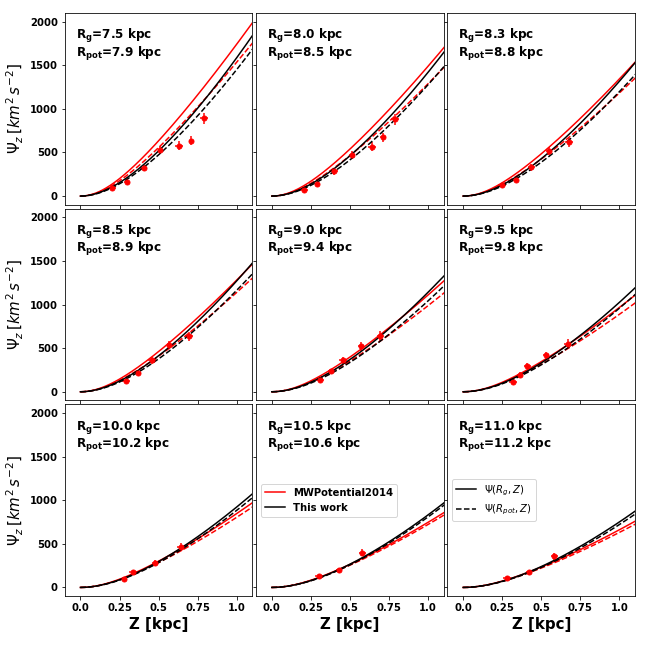}
\caption{
\label{fig:comp_global}
Vertical potential profiles from the global mass modelling with the interpolated snail intersections. The interpolated snail potential data points (red points) are fitted with a model consisting of a double exponential disc and an NFW dark matter halo (black lines), using the corresponding potential radii $R_{pot}$ of the phase snails. The MWPotential2014 from \cite{Bovy2015} is shown as the red lines for comparison. The solid lines show the potential at each $R_{g}$ bin center (as labelled), while the dashed lines display the potential at $R_{pot}$, corrected from the $R_{g}$ values using the relation in Fig. \ref{fig:obs_Rpot}.}
\end{figure*}

In the mass modeling, we choose the corrected potential radius for each phase snail to derive the vertical potential $\Psi(R_{pot}, z)$. Results of the global mass modeling of the interpolated snail intersections are shown in Fig. \ref{fig:comp_global}. For comparison, the potential profiles from MWPotential2014, and the profiles at the uncorrected radius (i.e. the center of each $R_{g}$ bin) $\Psi(R_{g}, z)$ from our best-fit potential model are also displayed. The model predicted vertical potential at $R_{pot}$, i.e. the black dashed lines, is well consistent with the data points. The vertical potential profiles at inner radii are consistent with those from the MWPotential2014 (red lines), for both $R_{g}$ and $R_{pot}$. At outer radii ($R_{g} > 10$ kpc), the snail intersections prefer steeper vertical potential, resulting in an overestimation of the outer vertical potential. Although we apply Gaussian priors for the disc scale lengths, the model predicted values are $4.22 \pm 0.27$ kpc and $3.99 \pm 0.57$ kpc, for the thin and thick discs, respectively. As our interpolated snail intersections cover the low vertical regions, our model gives a weak constraint on the thick disc, indicated by a peak of posterior distribution of $M_{d2}$ close to the lower boundary. Therefore, the overestimation of the vertical potential at outer radii indicates a shallower radial variation, i.e. a larger scale length of the thin disc than its prior.

As a summary, under strong priors on parameters of the mass model, we obtain a global Milky Way potential, roughly consistent with previous literature. The degeneracy between the different Milky Way components is still strong, requiring measurements in higher vertical regions to break the degeneracy. Nevertheless, we could measure the snail intersections for more radial bins with smaller intervals and overlapped stars. In such way, we could obtain a 2D vertical potential map in a radial range of $7\lesssim R \lesssim 11$ kpc. It thus might reduce the uncertainty in the mass modelling, or be directly utilized to give the vertical potential through interpolation.

%%%%%%%%%%%%%%%%%%%%%%%%%%%%%%%%%%%%%%%%%
%%%%%%%%%%%%%%%%%%%%%%%%%%%%%%%%%%%%%%%%%

%%%%%%%%%%%%%%%%%%%%%%%%%%%%%%%%%%%%%%%%%
\subsection{Modelling the Solar Neighborhood}
\label{ssec:mod_SN}
%%%%%%%%%%%%%%%%%%%%%%%%%%%%%%%%%%%%%%%%%
%%%%%%%%%%%%%%%%%%%%%%%%%%%%%%%%%%%%%%%%%
\subsubsection{Snail at $R_{pot}= R_{0}$}
\label{sssec:mod_SN_snail}
\begin{figure*}[htbp]
\centering
\includegraphics[width=0.9\textwidth]{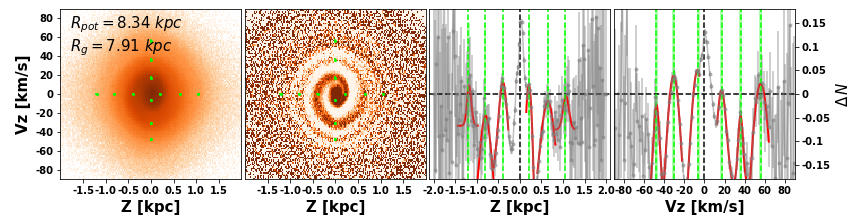}
\caption{
\label{fig:SN_snail}
Same as Fig. \ref{fig:obs_inter} but for the radial bin at $R_{g}= 7.91$ kpc. This phase snail is related to the potential radius of $R_{0}$. One additional outer wrap is measured compared to Fig. \ref{fig:obs_inter}.}
\end{figure*}

%%%%%%%%%%%%%%%%%%%%%%%%%%%%%%%%%%%%%%%%%
As mentioned in Section \ref{sec:method}, the snail at $R_{g}= 8.34$ kpc can not represent the vertical potential at the Solar neighborhood. According to the empirical relation shown in Fig. \ref{fig:obs_Rpot}, a potential radius of $R_{pot}= R_{0}= 8.34$ kpc is related to the snail at $R_{g}= 7.91$ kpc, showing a 0.43 kpc difference. Besides, according to the top three rows in Fig. \ref{fig:obs_inter}, snails seem to show one more wrap in the outer phase space. Therefore, in order to study the vertical potential from the snail intersections in the Solar neighborhood, we explore the snail within $|R_{g}-7.91|< 0.25$ kpc and measure the intersections for one more wrap than that in the previous Section. The measurement of the snail intersections is shown in Fig. \ref{fig:SN_snail}, which is similar to Fig. \ref{fig:obs_inter}. There are 12 snail intersections, resulting in 10 snail potential data points up to $z \sim 1.2$ kpc after the interpolation.

%%%%%%%%%%%%%%%%%%%%%%%%%%%%%%%%%%%%%%%%%
%%%%%%%%%%%%%%%%%%%%%%%%%%%%%%%%%%%%%%%%%

\begin{figure}[htbp]
\centering
\includegraphics[width=0.9\columnwidth]{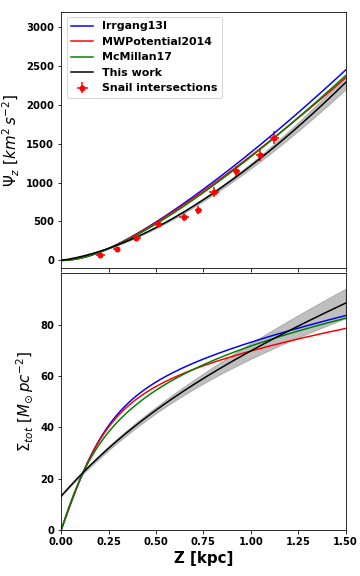}
\caption{
\label{fig:SN_inter}
Model predicted local vertical potential (top panel) and total surface density profile (bottom panel) by fitting the interpolated snail intersections (red points). The black lines show the median profiles derived from the posterior parameter sets, while the shaded regions indicate the $1\sigma$ errors. The colored lines display the vertical potential at the Solar radius from three popular Milky Way potentials for comparisons.}
\end{figure}

The interpolated snail intersections are shown in the upper panel of Fig. \ref{fig:SN_inter} as the red dots. For comparison, we also show the Solar vertical potential from three popular Milky Way potentials (i.e. Irrgang13I, MWPotential2014, McMillan17) as the colored lines. The snail intersections are well consistent with the three Milky Way potentials, except for two points at $z \sim 0.7$ kpc. These two points could suffer from the same problem as explained in Section \ref{ssec:obs_comp} for the snail at $R_{g}= 7.5$ kpc.
%The consistency would be better if the two points are slightly shifted to the left (i.e. smaller $Z_{max}$ with the same $V_{z,max}$ and $E_{z}$).

%%%%%%%%%%%%%%%%%%%%%%%%%%%%%%%%%%%%%%%%%
%%%%%%%%%%%%%%%%%%%%%%%%%%%%%%%%%%%%%%%%%

We apply a simple mass model to the snail intersections at $R_{pot}= R_{0}$, which consists of an exponential thin disc, a razor thin gaseous disc and a constant local dark matter density ($\rho_{\rm dm}$). This model is the same as that in \cite{GuoR2020}. The vertical potential is therefore written as:
%%%%%%%%%%%%%%%
\begin{equation}
\label{eq:sn_phiz}
\Psi(z)=2 \pi G\left\{\Sigma_{\star}\left[z-{\rm Z_{h}}+{\rm Z_{h}} \exp \left(-\frac{z}{\rm Z_{h}}\right)\right]+\Sigma_{\text {gas }} z+\rho_{\text {dm}} z^2\right\} \ ,
\end{equation}
%%%%%%%%%%%%%%%
where $\Sigma_{\star}$ and $\rm Z_{h}$ are the total surface density and the scale height of the stellar disc, respectively. The total gas surface density $\Sigma_{\rm gas}$ is set as 13.2 $\rm M_{\odot}\,pc^{-2}$ \citep{Flynn2006}. The free parameters left are $\mathbf{p}= (\Sigma_{\star},\ {\rm Z_{h}},\ \rho_{\rm dm})$. See the derivation of this potential model in Appendix \ref{sec:app_msn} or in \cite{GuoR2020}.

To consider the uncertainty in the $Z_{max}$ estimation, we perform 1000 Monte Carlo realizations according to the error of each intersection. The uncertainty of $\Psi(Z_{max})$ is absorbed into the following log-likelihood:
%%%%%%%%%%%%%%%
\begin{equation}
\label{eq:sn_lnlike}
\begin{aligned}
\ln L = &-\frac{1}{2} \sum_i\left[\frac{\left(\Psi_{\text{obs},i}-\Psi_{\text { model },i}(Z_{max,i})\right)^2}{\sigma_{\Psi, i}^2}\right] \\
   &-\sum_i \ln \left(\sqrt{2 \pi} \sigma_{\Psi, i}\right) \ ,
\end{aligned}
\end{equation}
%%%%%%%%%%%%%%%
where $\Psi_{{\rm obs},i}$ and $\sigma_{\Psi, i}$ are the interpolated vertical potential and its uncertainty for the $i$-th snail intersection, while $\Psi_{\text { model }, i}(Z_{max,i})$ is the model vertical potential at the related $Z_{max,i}$ for this intersection. We further apply Gaussian priors for $\Sigma_{\star}$ and $\rm Z_{h}$, which are $37.0 \pm 5.3\ {\rm M_{\odot}\,pc^{-2}}$ \citep{GuoR2020} and $0.5 \pm 0.1 $ kpc, respectively. For the local dark matter density, we just constrain it in a range of $[0, 0.05]$ $\rm M_{\odot}\,pc^{-3}$, with a flat prior.

\begin{figure}[htbp]
\centering
\includegraphics[width=0.9\columnwidth]{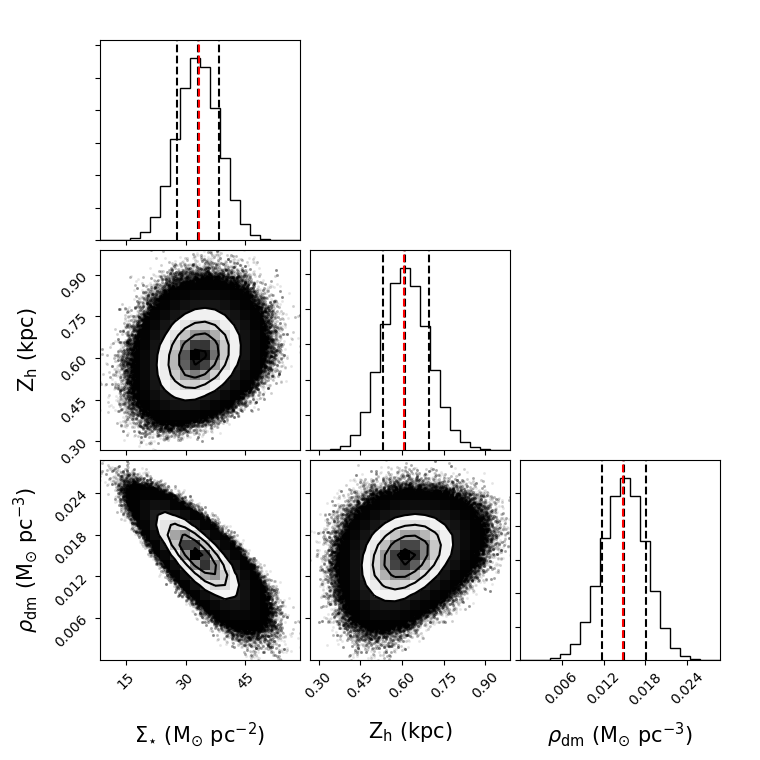}
\caption{
\label{fig:SN_coner}
Posterior probability distributions of parameters for the mass modelling of the snail at $R_{pot}= R_{0}$. The parameters from the left to right are the stellar surface density ($\Sigma_{\star}$), the stellar disc scale height (${\rm Z_{h}}$) and the local dark matter density ($\rho_{\rm dm}$), respectively. The black dashed lines in the histograms show the medians and $1\sigma$ errors of the marginalized 1D distributions, while the red dashed lines indicate the best-fit parameters.}
\end{figure}

The posterior distributions of these parameters from the MCMC method are shown in Fig. \ref{fig:SN_coner}. The medians and $1\sigma$ uncertainties of the 1D marginalized distributions are $33.1\pm5.2$ $\rm M_{\odot}\,pc^{-2}$, $0.612\pm0.085$ kpc, $0.0150\pm0.0031$ $\rm M_{\odot}\,pc^{-3}$ for the disc surface density, the disc scale height and the local dark matter density, respectively. The best-fit parameters are in coincidence with the medians, as shown by the dashed lines in the histogram panels in Fig. \ref{fig:SN_coner}.

The model predicted vertical potential profile and the corresponding total surface density profile are shown as the black solid lines in Fig. \ref{fig:SN_inter}, with the shaded regions representing their $1\sigma$ errors. The model predicted vertical potential is still systematically smaller than the three Milky Way potentials. This is mainly due to the large uncertainties of the two snail intersections at $z \sim 0.7$ kpc (resulting from the possible additional wrap), rather than due to the potential radius correction. For the total surface density, i.e. the slope of the vertical potential, the model predicted one shows significant difference compared to the three Milky Way potentials. There are two reasons. The first is that we set the constant gas surface density. This would influence the surface density profile at the low vertical region ($z <0.2$ kpc), as the scale height of the neutral hydrogen is about 0.085 kpc \citep{McMillan2017}. The second reason is the possible overestimation of $Z_{max}$ for the two snail intersections at $z \sim 0.7$ kpc, which will decrease the potential slope at the lower vertical height and increase the potential slope at the higher vertical height. The shape of the total surface density profile is then bent downward accordingly, which may explain the difference at $0.5< z < 1.0$ kpc.

%%%%%%%%%%%%%%%%%%%%%%%%%%%%%%%%%%%%%%%%%
%%%%%%%%%%%%%%%%%%%%%%%%%%%%%%%%%%%%%%%%%

%%%%%%%%%%%%%%%%%%%%%%%%%%%%%%%%%%%%%%%%%
\subsubsection{Comparison with previous works}
\label{sssec:mod_SN_comp}
\begin{figure}[htbp]
\centering
\includegraphics[width=0.9\columnwidth]{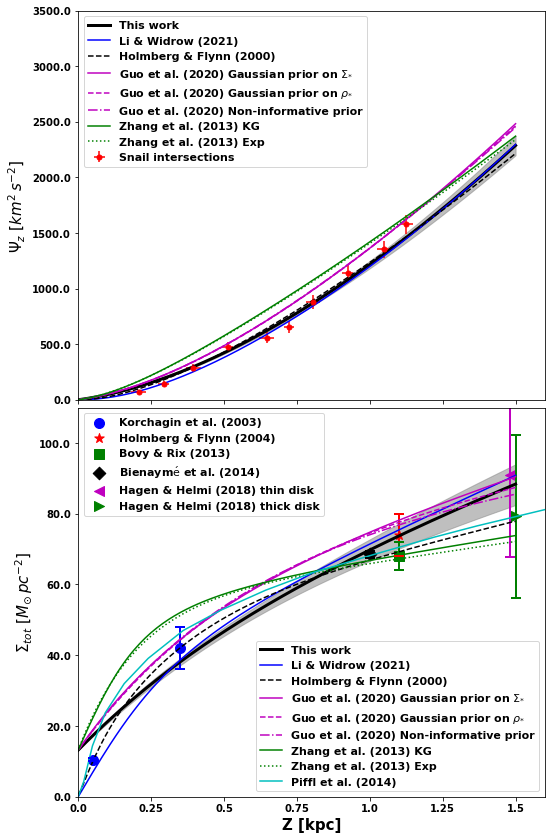}
\caption{
\label{fig:SN_liter}
Comparing the model predicted vertical potential (top panel) and the total surface density profile (bottom panel) at $R_{pot}= R_{0}$ with previous results in the literature (as labelled). The black lines show our modeling results by fitting the interpolated snail intersections (red points), with shaded regions indicating the $1\sigma$ errors. Our results best agree with the profiles from \cite{Holmberg2000} (black dashed lines) and \cite{LiHC2021} (blue solid lines).}
\end{figure}

%%%%%%%%%%%%%%%%%%%%%%%%%%%%%%%%%%%%%%%%%
There are a lot of works trying to measure the vertical potential and mass distribution in the Solar neighborhood \citep[e.g.][]{ZhangL2013, XiaQR2016, Hagen2018, Schutz2018, Sivertsson2018, Buch2019, GuoR2020, Salomon2020}. These works utilize dynamical methods, such as the distribution function and the vertical Jeans equation, to model the stellar kinematics, based on the equilibrium state assumption. Here, we compare the vertical potential profile and the total surface density profile of our independent measurement with those from previous works, e.g. \cite{Holmberg2000, Korchagin2003, Holmberg2004, Bovy2013, ZhangL2013, Bienayme2014, Piffl2014, GuoR2020, LiHC2021}, as shown in the upper and lower panels of Fig. \ref{fig:SN_liter}, respectively. The red points and the black solid lines are the same as those in Fig. \ref{fig:SN_inter}.

As shown in the upper panel of Fig. \ref{fig:SN_liter}, our model predicted vertical potential profile, i.e. the median of all potential profiles derived from the posterior parameter sets, shows quite good agreement with the profiles from \cite{LiHC2021} and \cite{Holmberg2000}. However, considering the possible overestimation of $Z_{max}$ at $z \sim 0.7$ kpc, the interpolated snail intersections would be more similar to our previous work \citep{GuoR2020}. The profiles from \cite{ZhangL2013} seem to slightly overestimate the vertical potential at $z \lesssim 1$ kpc, relative to other works. In the lower panel, if ignoring the constant gas surface density applied, our surface density profile is best consistent with that from \cite{LiHC2021}. However, as explained previously, the shape in the region of about $0.5< z <1.0$ kpc should be bent upward slightly. Thus the intrinsic $\Sigma_{\rm tot}$ profile might be more similar to our previous work \citep{GuoR2020}. The profiles from \cite{ZhangL2013} seem to be slightly larger in the lower region ($z<1$ kpc) and smaller in the higher region ($z>1$ kpc), where \cite{Piffl2014} also shows a shallower profile. Taking the constant $\Sigma_{\rm gas}$ and the probable overestimation of $Z_{max}$ at $z \sim 0.7$ kpc into consideration, our model predicted surface density profile could be consistent with all the measurements shown by the different symbols in the lower panel of Fig. \ref{fig:SN_liter}.

For the local dark matter density, our model predicts $\rho_{\rm dm}= 0.0150\pm0.0031$ $\rm M_{\odot}\,pc^{-3}$, which is consistent with our previous measurement of $0.0133_{-0.0022}^{+0.0024}\ {\rm M}_{\odot }\, {\rm pc}^{-3}$ \citep{GuoR2020} and other works within the error bars. It is larger than some other works \citep{Read2014, deSalas2021}. However, if we consider the probable overestimation of $Z_{max}$ at $z \sim 0.7$ kpc, the $\Sigma_{\rm tot}$ profile should be bent upward in the region of $0.5<z<1.0$ kpc. It means that the total surface density will increase more quickly in the lower region, and more slowly in the higher region. Therefore, the scale height of the stellar disc should be smaller, which is also more reasonable as the model predicted $\rm Z_{h}= 0.612\pm0.085$ kpc is larger for the thin disc \citep{McKee2015}. The local dark matter density should be slightly smaller considering the smaller slope of $\Sigma_{\rm tot}$ in the higher region, where the constant dark matter dominates the profile.

%%%%%%%%%%%%%%%%%%%%%%%%%%%%%%%%%%%%%%%%%
%%%%%%%%%%%%%%%%%%%%%%%%%%%%%%%%%%%%%%%%%

%%%%%%%%%%%%%%%%%%%%%%%%%%%%%%%%%%%%%%%%%%%%%%%%%%%%%%%%%%%%%%
\section{Discussions}
\label{sec:discuss}
%%%%%%%%%%%%%%%%%%%%%%%%%%%%%%%%%%%%%%%%%%%%%%%%%%%%%%%%%%%%%%
%%%%%%%%%%%%%%%%%%%%%%%%%%%%%%%%%%%%%%%%%
\subsection{Comparison with the phase snail modeling in previous works}
\label{ssec:comp_liter}
%%%%%%%%%%%%%%%%%%%%%%%%%%%%%%%%%%%%%%%%%
Since the discovery of the phase snail, many works have tried to measure the Milky Way mass distribution and potential with the phase snail. In the series works of \cite{Widmark2021a, Widmark2021b, Widmark2022a, Widmark2022b}, the authors constructed the analytical snail model through the 1D phase-space orbital integration and a sinusoidal function of the action angle. Assuming a constant amplitude relative to the background, they obtained an estimation of the vertical potential embedded in the phase-space orbital integration by fitting the 2D phase space distribution in the observation. In their model, the orbital integration requires a potential model and the time since the perturbation. In two other works, \cite{LiHC2021, LiHC2022} simultaneously modeled the gravitational potential and phase space distribution function of giant stars under the steady-state assumption of the disk, with clear snail features shown in their residual maps. In our previous work \citep{GuoR2022}, we applied a geometric snail model superimposed on a smooth background in the phase space to fit the north-south asymmetric vertical velocity dispersion profile. The phase space distribution of the smooth background component is built through the vertical Jeans equation and the background potential, while the snail model is constructed empirically from both the observation and test particle simulations. In a different work, \cite{Price-Whelan2021} utilized Orbital Torus Imaging, which makes use of kinematic measurements and element abundances, to investigate the metallicity distribution in the phase space. Under the steady-state, stellar labels (e.g. metallicity) should vary systematically with orbit characteristics such as actions, while keeping invariant with respect to the action angle ($\theta_{z}$). By minimizing the residual metal distribution as a function of $\theta_{z}$, they also obtained an estimation of the Milky Way mass model.

There are many works on the phase snail focusing on the measurement of the time since the perturbation, which is found to vary with the angular momentum \citep{Antoja2022, Darragh-Ford2023}. \cite{Antoja2022} measured the snail shapes first, which vary with the angular momentum and the azimuth significantly. They then inferred the time since the perturbation by the difference of vertical frequencies of snail intersections. The derived impact time varies with both the angular momentum and the vertical height. In their work, a Milky Way potential model is needed to derive the vertical frequencies of snail intersections, though different potential models lead to similar results. In \cite{Darragh-Ford2023}, the time since the perturbation is measured from the slope of the action angle ($\theta_{z}$) as a function of the vertical frequency ($\Omega_{z}$). They also need a potential model to derive $\theta_{z}$ and $\Omega_{z}$ of each star, which are also utilized to extract the snail shape. In order to find a suitable potential, they ran the action calculations in a grid of potential parameter values, and visually picked a potential model in which the snail is close to a straight line in the space of $\sqrt{J_{z}}-\theta_{z}$.

In our work, the toy model snails in Section \ref{ssec:method_test} are built through the 1D orbital integration similar to \cite{Widmark2021a}. However, these toy model snails are only used to distinguish the difference between $R$- and $R_{g}$-binned snails and to refine the interpolation method. Different from the previous works, the linear interpolation method here does not need to assume any potential model. In addition, although the guiding radius calculation is based on a rotation curve, which is an implication of the underlying mass distribution, we can actually separate the sample according to the angular momentum, which is directly derived from the observation. In a summary, our method provides a new way to directly obtain potential measurements, independent from the mass modeling and the assumption of the time since the perturbation.

%%%%%%%%%%%%%%%%%%%%%%%%%%%%%%%%%%%%%%%%%
\subsection{Connection between the phase snail and the North/South asymmetry}
\label{ssec:obs_AzAvz}
\begin{figure}[htbp]
\centering
\includegraphics[width=0.8\columnwidth]{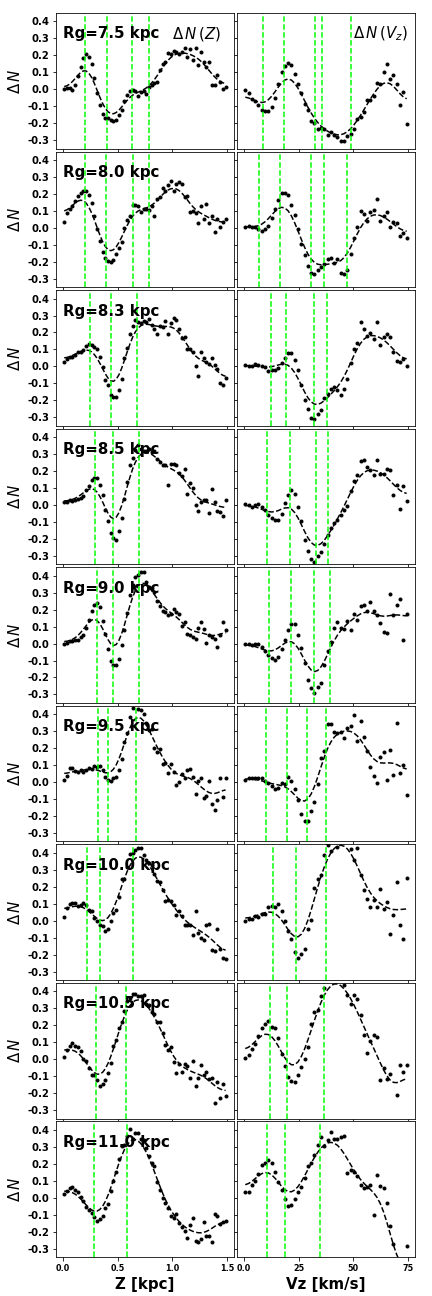}
\caption{
\label{fig:obs_AzAvz}
The number density difference between the positive and negative vertical height (left) and vertical velocity (right) at different guiding center radii (as labelled in each row). The black dashed lines are the Gaussian smoothing profiles of the data points. The light green vertical lines indicate the snail intersections measured in Fig. \ref{fig:obs_inter}.}
\end{figure}

%%%%%%%%%%%%%%%%%%%%%%%%%%%%%%%%%%%%%%%%%
In our previous work \citep{GuoR2022}, we found that the north and south asymmetry of the Galactic plane can be quantitatively explained by a simple empirical snail model. The number density difference profile between north and south, i.e. $A_{z}= (\nu(z)-\nu(-z))/(\nu(z)+\nu(-z))$, shows a wave-like fluctuation \citep{Widrow2012, Bennett2019}, which is caused by the snail intersecting with the $z$ axis at different vertical heights. To inspect this effect and check if it could be utilized to measure the snail intersections, we investigate the number density difference profiles between the positive and negative vertical height ($A_{z}$) and vertical velocity ($A_{V_{z}}= (\nu(V_{z})-\nu(-V_{z}))/(\nu(V_{z})+\nu(-V_{z}))$) at different $R_{g}$. We select the same stripes along $z$ and $V_{z}$ axes as Fig. \ref{fig:obs_inter} to check the number density difference profiles, in order to avoid the cumulative effect of different snail wraps when using the whole phase space. Note that, we ignore the selection effects and regard they are the same at positive and negative $z$ or $V_{z}$. The results are shown as the black dots in Fig. \ref{fig:obs_AzAvz}. The snail intersections measured in the previous section are indicated by the vertical dashed lines. We also show Gaussian smoothed profiles for the data points, shown as the black dashed lines in Fig. \ref{fig:obs_AzAvz}.

For most snail intersections, they are consistent with the locations of peaks and troughs, especially at smaller vertical heights and velocities. This is due to the higher number density for regions closer to the phase space origin. At larger vertical heights and velocities, due to the large numerical noise, the peaks and troughs are sometimes not obvious. It results in a weak correlation with the snail intersections. Besides, the snail wraps are tighter at inner radii, which results in smaller absolute difference between the adjacent intersections along $z$ and $V_{z}$ axes. Therefore, the adjacent peaks and troughs of $A_{z}$ and $A_{V_{z}}$ profiles could be mixed up, as indicated by the outer two intersections in the top left panel of Fig. \ref{fig:obs_AzAvz}. Another interesting result is the amplitudes of the $A_{z}$ and $A_{V_{z}}$ profiles, which vary with the guiding center radii, the vertical heights and vertical velocities. The amplitudes may include the information about the perturbation strength, which will be investigated in our future work.

%%%%%%%%%%%%%%%%%%%%%%%%%%%%%%%%%%%%%%%%%
%%%%%%%%%%%%%%%%%%%%%%%%%%%%%%%%%%%%%%%%%

%%%%%%%%%%%%%%%%%%%%%%%%%%%%%%%%%%%%%%%%%
\subsection{Some caveats in our method}
\label{ssec:caveats}
%%%%%%%%%%%%%%%%%%%%%%%%%%%%%%%%%%%%%%%%%
One important assumption in our work is that the vertical motion is separable from the in-plane motions and the vertical energy is conserved, which thus relates $Z_{max}$ and $V_{z,max}$ of snail intersections as Eq. \ref{eq:Ez_zvzmax}. This is an appropriate assumption at lower vertical heights, but breaks down at high vertical regions. As shown in Fig. \ref{fig:snail_diff}, our toy model snail from 1D vertical integration can roughly trace the ridge line of the snail in the test particle simulation. At the outer region of phase space (about $z> 1.5$ kpc), the snails always stay inside the toy model snails, due to the ignorance of the in-plane motions. To trace the vertical potential from snails at large vertical heights, 3D orbit integration is needed. Otherwise, it would result in a shallower profile at high regions.

Another caveat is the empirical potential radius correction for $R_{g}$-binned snails. Though our empirical correction method is verified with the test particle simulation, it should be carefully applied in the observations. Both the sample selection and the data binning may have slight influence on the resultant correction relation. The decline at the tail in Fig. \ref{fig:obs_Rpot} could be due to the incompleteness at large radii. Besides, stars in cold orbits and hot orbits should have different correction relations, as their radial actions are different. As illustrated in \cite{LiZY2020}, the hotter orbits phase-mix more quickly and may show no snail features in the phase space. We select all stars in this work, which results in a $0.4-0.5$ kpc difference between $R_{g}$ and $R_{pot}$. If only stars with smaller radial actions are selected, this difference could be smaller. The accurate correction is difficult to perform for each star, which may need the 3D orbit integration. More detailed studies are needed to accurately quantify the connection between the shapes of $R_{g}$ snails and the underlying vertical potential.

%%%%%%%%%%%%%%%%%%%%%%%%%%%%%%%%%%%%%%%%%
\subsection{The self-gravity effect}
\label{ssec:self-g}
%%%%%%%%%%%%%%%%%%%%%%%%%%%%%%%%%%%%%%%%%
The self-gravity is ignored in our work, especially when testing our method with the test particle simulation and building the model snails. However, the self-gravity is important in modelling the evolution of the snails, especially the tightness of the snails. The phase snails found in N-body simulations are not as tightly wound as in the observations \citep{Laporte2019, Bennett2022}. \cite{Darling2019} compared N-body simulations of a test particle disc with fully self-consistent ones. They found that the snails in the simulations with the self-gravity are less tightly wound and more difficult to discern. More recently, \cite{Widrow2023} explored the influence of the self-gravity on the amplitude and pitch angle of the phase snails within the framework of the shearing box approximation. With the self-gravity, the surface density is amplified due to the swing amplification, which thus sets off new vertical phase snails. The snails are then less tightly wound while stronger than those in the case without the self-gravity. The slope in the frequency and action angle space therefore indicates a smaller perturbation time, related to the time between the swing amplification peak and the snapshot time.

Our interpolation method bases on the linear relations between quantities and the vertical frequency. When the self-gravity is considered, the vertical frequencies of snail intersections are gradually changed, with different extent. Though the phase angle difference between adjacent intersections is still $\pi/2$ and thus the vertical frequency difference keeps constant, the approximate linear relations between quantities and $\Omega_{z}$ might be changed. On the other hand, these relations are generally accurate for different vertical potential profiles at different radii. As a balance, it is difficult to assess if the interpolation of averaging the adjacent intersections would be problematic or not. In addition, the self-gravity would also gradually change the vertical density profile and thus the vertical potential profile. The shape of the phase snail contains both the information of the vertical potential profile before the perturbation and its gradual variation due to the self-gravity. The vertical potential traced by the snail seems neither to be the profile before the perturbation, nor the current profile. Finally, the strength of the perturbation, i.e. the fraction of stars being strongly perturbed or in the snail structure, is important in assessing the influence of the self-gravity. Studies on the amplitude of the phase snail \citep[e.g.][]{GuoR2022, Alinder2023, Frankel2022} are helpful to understand it. All these considerations need more detailed studies with the self-gravity considered. We leave them to the future work.

%%%%%%%%%%%%%%%%%%%%%%%%%%%%%%%%%%%%%%%%%%%%%%%%%%%%%%%%%%%%%%
\section{Conclusion}
\label{sec:con}
%%%%%%%%%%%%%%%%%%%%%%%%%%%%%%%%%%%%%%%%%%%%%%%%%%%%%%%%%%%%%%
The vertical phase snail is a direct sign of dis-equilibrium of our Milky Way, which is the result of incomplete phase-mixing. Nevertheless, the snail wraps mainly under the vertical potential, which means it contains the information of the Galactic mass distribution of different components. In this work, we propose a novel method to measure the Galactic vertical potential, utilizing the intersections between the phase snails and $z/V_{z}$ axes (i.e. the turning-around and mid-plane points). The vertical motion is assumed to be decoupled with the in-plane motion, which is a proper assumption at low vertical height ($z< 1.5$ kpc). The snail intersections have known maximum vertical heights ($Z_{max}$) for the turning-around points or vertical velocities ($V_{z,max}$) for the mid-plane ones. Applying an interpolation method, we can thus obtain ($Z_{max}$, $\frac{1}{2} V_{z,max}^{2}$) for each intersection, which then provides a direct and model independent measurement of the vertical potential.

This method has been tested utilizing a test particle simulation. We find approximate linear relations of $V_{z,max}$ and $\sqrt{Z_{max}}$ with the vertical frequency $\Omega_{z}$. Thus, they can be utilized in the linear interpolation between the adjacent snail intersections, which have a constant action angle interval $\theta{z}= \pi/2$ and thus a constant $\Delta \Omega_{z}$. In addition, we find that the snail binned by the guiding center radius is quantitatively different from that binned by the Galactocentric radius. Direct shape comparison shows that the $R_{g}$-binned snail stays slightly inside the $R$-binned snail, which means it stands for a vertical potential shallower than that for the $R$-binned snail. This is due to the fact that stars on a non-circular orbit stay longer outside the guiding center radius. Thus on average, the $R_{g}$-binned snail stands for a vertical potential with a potential radius larger than $R_{g}$, i.e. $R_{pot} > R_{g}$. This difference can also be understood as the asymmetric drift: the median $R_{g}$ of a sample in a given radial range is smaller than the center of the radial bin, i.e. $R_{pot}$. The difference between $R_{pot}$ and $R_{g}$ depends on the radius, the radial action of the tracer and thus the sample selection. We apply an empirical method to relate $R_{g}$ with $R_{pot}$. From our test particle simulation, the difference between $R_{pot}$ and $R_{g}$ is about 0.1-0.15 kpc, which results in about 5\% systematic underestimation of the vertical potential with $R_{g}$ not corrected. After the correction from $R_{g}$ to $R_{pot}$, our interpolation method obtains an estimation of the vertical potential with less than 5\% random errors.

We apply our method to the Gaia DR3 for several $R_{g}$ bins. The empirical relation between the potential radius and guiding center radius for our sample shows that $R_{g}$ is about 0.4 kpc smaller than $R_{pot}$, which needs to be corrected in the potential comparison and the further mass modeling. We apply a Gaussian kernel convolution to obtain the density contrast profiles for stripes along $z/V_{z}$ axes. The snail intersections are measured by Gaussian fitting to visually selected regions where peaks are located. With the measured $Z_{max}$ for the turning-around points and $V_{z,max}$ for the mid-plane points, we apply the linear interpolation method as refined in Eq. \ref{eq:interp} to obtain $(Z_{max},\ \frac{1}{2}V_{z,max}^{2})$ for these intersections. Note that, at $R_{g}= 7.5, 8.0$ kpc, the maximum vertical heights of the intersections at $z \sim 0.7$ kpc seem to be overestimated, resulting in overestimation of $Z_{max}$ for adjacent snail intersections. In addition, the locations of intersections are also coincident with some peaks and troughs of the difference profiles between the positive and negative vertical heights ($A_{z}$) and vertical velocities ($A_{V_{z}}$), which might be utilized to locate the snail intersections.

We compare the potential measurements for the snail intersections with three popular Milky Way potentials, i.e. the Model I from \cite{Irrgang2013}, the MWPotential2014 from \cite{Bovy2015} and the McMillan17 from \cite{McMillan2017}. Our measurements are more consistent with the vertical potential at the potential radii, which are obviously shallower than the vertical potential at the smaller guiding center radii. At larger radii, the difference is smaller as the radial variation of the vertical potential is smaller. We apply further mass modeling to the snail intersections at different radii, with the scale heights fixed and priors applied to the scale lengths of the double exponential disc. The resultant vertical potential is well fitted to the snail intersections and consistent with the potential from \cite{Bovy2015} at small radii. However, it shows an overestimation at large radii.

From the empirical relation between $R_{g}$ and $R_{pot}$ for our sample, we attempt to measure the vertical potential in the Solar neighborhood, which is related to the snail at $R_{g}= 7.91$ kpc. We manage to measure the snail intersections for one more wrap with relatively large uncertainties. The resultant potential measurements for the snail intersections are consistent with the Solar vertical potential from the three popular Milky Way potentials. We fit the interpolated vertical potential of these intersections with a mass model consisting of a thin stellar disc, a razor thin gaseous disc and a constant local dark matter density. Under the priors on the stellar total surface density and the disc's scale height, we obtain a local dark matter density of $\rho_{\rm dm}= 0.0150\pm0.0031$ $\rm M_{\odot}\,pc^{-3}$, which is consistent with previous works. The model predicted vertical potential profile and the vertical total surface density profile are consistent with previous results, especially with \cite{Holmberg2000} and \cite{LiHC2021}. However, if we take the possible overestimation of $Z_{max}$ at $z \sim 0.7$ kpc into consideration, the intrinsic vertical potential could be more consistent with \cite{GuoR2020}. The profile of $\Sigma_{\rm tot}$ would be bent upwards, resulting in a smaller stellar vertical scale height and a smaller local dark matter density.

In principle, this model independent method can be applied to more radial bins with overlapped stars. In this way, we can better compare with different known Milky Way potentials, and provide a stronger constraint on the Milky Way potential. Our work indicates that the vertical phase snail can indeed provide a new way to study the Milky Way's vertical potential. In our future work, we will try to measure the vertical potential utilizing the whole snail shape curve, rather than just the snail intersections, which might also better constrain the Milky Way potential.

\acknowledgments
We thank the anonymous reviewers for their thoughtful comments.
RG is supported by Initiative Postdocs Supporting Program (No. BX2021183), funded by China Postdoctoral Science Foundation. The research presented here is partially supported by the National Key R\&D Program of China under grant No. 2018YFA0404501; by the National Natural Science Foundation of China under grant Nos. 12103031, 11773052, 11761131016, 11333003, 12025302, 12122301, 12233001; and by the ``111'' Project of the Ministry of Education under grant No. B20019 and by the Shanghai Natural Science Research Grant (No. 21ZR1430600). We acknowledge the science research grants from the China Manned Space Project with No. CMS-CSST-2021-B03. J.S. acknowledges support from a Newton Advanced Fellowship awarded by the Royal Society and the Newton Fund. ZYL acknowledges the sponsorship from Yangyang Development Fund. This project was developed in part at the LAMOST-Gaia Sprint 2022, supported by the National Natural Science Foundation of China (NSFC) under grants 11873034 and U2031202.
 
This work made use of the Gravity Supercomputer at the Department of Astronomy, Shanghai Jiao Tong University, and the facilities of the Center for High Performance Computing at Shanghai Astronomical Observatory.
%Guoshoujing Telescope (the Large Sky Area Multi-Object Fiber Spectroscopic Telescope LAMOST) is a National Major Scientific Project built by the Chinese Academy of Sciences. Funding for the project has been provided by the National Development and Reform Commission. LAMOST is operated and managed by the National Astronomical Observatories, Chinese Academy of Sciences. 
This work has made use of data from the European Space Agency (ESA) mission {\it Gaia} (\url{https://www.cosmos.esa.int/gaia}), processed by the {\it Gaia} Data Processing and Analysis Consortium (DPAC, \url{https://www.cosmos.esa.int/web/gaia/dpac/consortium}). Funding for the DPAC has been provided by national institutions, in particular the institutions participating in the {\it Gaia} Multilateral Agreement.


\vspace{0.4\textwidth}
%\newpage

%%%%%%%%%%%%%%%%%%%%%%%%%%%%%%%%%%%%%%%%%%%%%%%%%%%%%%%%%%%%%%
%\begin{appendices}
\appendix
\twocolumngrid
%%%%%%%%%%%%%%%%%%%%%%%%%%%%%%%%%%%%%%%%%%%%%%%%%%%%%%%%%%%%%%
\section{Mass Model for Global Modelling}
\label{sec:app_mglobal}
%%%%%%%%%%%%%%%%%%%%%%%%%%%%%%%%%%%%%%%%%%%%%%%%%%%%%%%%%%%%%%
The potential $\Phi(R,z)$ of a galaxy is related to its mass distribution by the Poisson equation:
%%%%%%%%%%%%%%%%%%%%%%%%%%%%%%%%%%%%%%
\begin{equation}
\label{eq:a_poisson}
\nabla^2 \Phi=4 \pi G \rho =\frac{1}{R} \frac{\partial}{\partial R}\left(R \frac{\partial \Phi}{\partial R}\right)+\frac{\partial^2 \Phi}{\partial z^2} \ .
\end{equation}
%%%%%%%%%%%%%%%%%%%%%%%%%%%%%%%%%%%%%%
For a flattened system, the radial derivative term is small compared to the vertical derivative term. Thus, \ref{eq:a_poisson} can be simplified to:
%%%%%%%%%%%%%%%%%%%%%%%%%%%%%%%%%%%%%%
\begin{equation}
\label{eq:a_ps}
\frac{\partial^2 \Phi (R,z)}{\partial z^2}=4 \pi G \rho (R,z) \ .
\end{equation}
%%%%%%%%%%%%%%%%%%%%%%%%%%%%%%%%%%%%%%
This form can be applied to almost any thin disk system \citep{Binney2008}, which is also equal to assuming a flat rotation curve ($V_{c}$) as $V_{c}^{2}= R\frac{\partial \Phi}{\partial R}$.

%%%%%%%%%%%%%%%%%%%%%%%%%%%%%%%%%%%%%%
For modeling of the global disc in Section \ref{ssec:mod_global}, we consider a mass model consisting of double exponential discs and an NFW \citep{NFW1996} dark matter halo. The double discs both have a density profile as:
%%%%%%%%%%%%%%%%%%%%%%%%%%%%%%%%%%%%%%
\begin{equation}
\label{eq:gm_rhodisc}
\rho_d(R, z) =\rho_{d, 0} \exp \left(-\frac{R}{R_d}\right) \exp \left(-\frac{z}{z_d}\right) = \nu_0 \exp \left(-\frac{z}{z_d}\right) \ ,
\end{equation}
%%%%%%%%%%%%%%%%%%%%%%%%%%%%%%%%%%%%%%
where $R_d$ and $z_d$ are the scale length and scale height of the disc, and $\rho_{d, 0}$ is a scale density. $\nu_0$ contains the radial variation of the density profile, related to the surface density profile as $\nu_0= \frac{\Sigma(R)}{2z_{d}}$. The total mass of the disc $M_d$ is thus
%%%%%%%%%%%%%%%%%%%%%%%%%%%%%%%%%%%%%%
\begin{equation}
\label{eq:gm_mdisc}
\begin{aligned}
M_d & =\int_0^{\infty} \rho_{d, 0} \exp \left(-\frac{R}{R_d}\right) \cdot 2 \pi R d R \cdot 2 \int_0^{\infty} \exp \left(-\frac{z}{z_d}\right) d z \\
& =4 \pi \rho_{d, 0} z_d \cdot \int_0^{\infty} \exp \left(-\frac{R}{R_d}\right) \cdot R d R \\
& =4 \pi \rho_{d, 0} z_d R_d^2 \ .
\end{aligned}
\end{equation}
%%%%%%%%%%%%%%%%%%%%%%%%%%%%%%%%%%%%%%

The potential of the exponential disc is complicated to be analytically formulated. It is usually related with some Bessel integral formula. In this work, for simplicity, we just integrate Eq. \ref{eq:a_ps} along $z$ direction to directly obtain the reduced vertical potential for both the thin and thick discs. Thus, the reduced vertical potential of a stellar disc is written as: 
%%%%%%%%%%%%%%%%%%%%%%%%%%%%%%%%%%%%%%
\begin{equation}
\label{eq:gm_phidisc}
\begin{aligned}
\Psi_{d} (z) & =\iint_0^z \nu_0 \exp \left(-\frac{z^{\prime}}{z_d}\right) d z^{\prime} \cdot 4 \pi G \\
& =4 \pi G \nu_0 z_d \int_0^z\left[1-\exp \left(-\frac{z^{\prime}}{z_d}\right)\right] d z^{\prime} \\
& =4 \pi G \nu_0 z_d \cdot\left[z-z_d+z_d \exp \left(-\frac{z}{z_d}\right)\right] \\
& =\frac{GM_{d}}{R_{d}^2} \exp \left(-\frac{R}{R_{d}}\right)\left[z-z_{d}+z_{d} \exp \left(-\frac{z}{z_{d}}\right)\right] \ .
\end{aligned}
\end{equation}
%%%%%%%%%%%%%%%%%%%%%%%%%%%%%%%%%%%%%%

For an NFW dark matter halo, the potential $\Phi_h (r)$ is:
%%%%%%%%%%%%%%%%%%%%%%%%%%%%%%%%%%%%%%
\begin{equation}
\label{eq:gm_phidm}
\Phi_h(r) =-4 \pi G \rho_{h, 0} a_h^2 \cdot \frac{\ln (1+r / a_h)}{r / a_h}=-G M_{h} \cdot \frac{\ln (1+r / a_h)}{r} \ ,
\end{equation}
%%%%%%%%%%%%%%%%%%%%%%%%%%%%%%%%%%%%%%
where $r=\sqrt{R^2+z^2}$. $\rho_{h, 0}$, $a_h$ and $M_h$ are the scale density, scale length and scale mass, respectively. The reduced vertical potential at ($R, z$) is then
%%%%%%%%%%%%%%%%%%%%%%%%%%%%%%%%%%%%%%
\begin{equation}
\label{eq:gm_phidm0}
\begin{aligned}
\Psi_h(R, z) & =\Phi_h(R, z)-\Phi_h(R, 0) \\
& =G M_h \cdot\left[\frac{\ln (1+R / a_h)}{R}-\frac{\ln (1+r / a_h)}{r}\right] \ .
\end{aligned}
\end{equation}
%%%%%%%%%%%%%%%%%%%%%%%%%%%%%%%%%%%%%%
Combining Eqs. \ref{eq:gm_phidisc} and \ref{eq:gm_phidm0}, we can obtain Eq. \ref{eq:gol_dexp}, i.e. the reduced vertical potential for a mass model consisting of double exponential discs and an NFW dark matter halo.

%%%%%%%%%%%%%%%%%%%%%%%%%%%%%%%%%%%%%%%%%%%%%%%%%%%%%%%%%%%%%%
\section{Mass Model for the Solar Neighborhood}
\label{sec:app_msn}
%%%%%%%%%%%%%%%%%%%%%%%%%%%%%%%%%%%%%%%%%%%%%%%%%%%%%%%%%%%%%%
For the Solar neighborhood, we only consider a simple mass model similar to \cite{GuoR2020}, which consists of an exponential thin disc, a razor thin gaseous disc and a constant local dark matter density ($\rho_{\text {dm}}$). The derivation of the reduced vertical potential for this model is clearly shown in \cite{GuoR2020} and is summarized here. The total volume density profile $\rho_{\text {tot}}(z)$ is:
%%%%%%%%%%%%%%%%%%%%%%%%%%%%%%%%%%%%%%
\begin{equation}
\label{eq:sn_rhot}
\rho_{\text {tot }}(z)  =\rho_{*, 0} \exp \left(-\frac{z}{\rm Z_h}\right)+\Sigma_{\text {gas }} \delta(z)+\rho_{\text {dm }} \ ,
\end{equation}
%%%%%%%%%%%%%%%%%%%%%%%%%%%%%%%%%%%%%%
where $\rho_{*, 0}$ is the stellar local volume density and $\rm Z_h$ is the stellar scale height. $\Sigma_{\text {gas }}$ is the total gaseous surface density, which is fixed as 13.2 $\rm M_{\odot}\,pc^{-2}$ \citep{Flynn2006} in this work.

Substituting this mass profile into Eq. \ref{eq:a_ps} and integrating it along $z$ direction at the Solar radius $R_{0}$, we obtain the vertical force $K_z(z)$, i.e. the first derivative of the vertical potential. It can also be scaled to the total surface density $\Sigma_{\text {tot }}(z)$ as $K_z(z)= -2 \pi G \Sigma_{\text {tot }}(z)$. Thus, the vertical force can be written as
%%%%%%%%%%%%%%%%%%%%%%%%%%%%%%%%%%%%%%
\begin{equation}
\label{eq:sn_kz}
\begin{aligned}
K_z(z) & \equiv -\frac{d \Phi}{d z}=-\int_0^z 4 \pi G \rho_{\text {tot }}\left(z^{\prime}\right) d z^{\prime}=-2 \pi G \Sigma_{\text {tot }}(z) \\
& =-2 \pi G\left\{\Sigma_*\left[1-\exp \left(-\frac{z}{\rm Z_h}\right)\right]+\Sigma_{\text {gas }}+2 \rho_{\text {dm }} z\right\} \ ,
\end{aligned}
\end{equation}
%%%%%%%%%%%%%%%%%%%%%%%%%%%%%%%%%%%%%%
where $\Sigma_*$ is the total stellar surface density. The reduced vertical potential of this mass model is therefore:
%%%%%%%%%%%%%%%%%%%%%%%%%%%%%%%%%%%%%%
\begin{equation}
\label{eq:sn_phiz}
\begin{aligned}
\Psi(z) & = \Phi(R_{0},z)-\Phi(R_{0},0) =-\int_0^z K_z\left(z^{\prime}\right) d z^{\prime} \\
& =2 \pi G\left\{\Sigma_*\left[z-{\rm Z_h}+{\rm Z_h} \exp \left(-\frac{z}{\rm Z_h}\right)\right]+\Sigma_{\text {gas }} z+\rho_{\text {dm }} z^2\right\}.
\end{aligned}
\end{equation}
%%%%%%%%%%%%%%%%%%%%%%%%%%%%%%%%%%%%%%

%\end{appendices}

%%%%%%%%%%%%%%%%%%%%%%%%%%%%%%%%%%%%%%%%%%%%%%%%%%%%%%%%%%%%%%



\end{document}